\documentclass[preprint]{aastex}

\begin{document} 

\title{[OII] emission, eigenvector~1 and orientation in radio-quiet quasars}

\author{J. Kuraszkiewicz\altaffilmark{1}, B. J. Wilkes} 
\affil{Harvard-Smithsonian Center for Astrophysics, Cambridge, MA
02138}

\author{W. N. Brandt}
\affil{Department of Astronomy and Astrophysics, 
The Pennsylvania State University, University Park, PA 16802}

\and

\author{M. Vestergaard}
\affil{The Ohio State University, Columbus, OH 43120\altaffilmark{2}}

\altaffiltext{1}{also N. Copernicus Astronomical Center, Warsaw,
Poland}
\altaffiltext{2}{also Harvard-Smithsonian Center for Astrophysics, 
Cambridge, MA 02138}

\begin{abstract}

We present supportive evidence that the Boroson \& Green eigenvector~1
is not driven by source orientation and further that both
[OIII]\,$\lambda$5007 and [OII]\,$\lambda3727$ are isotropically
emitted in the radio-quiet sample of BQS (Bright Quasar Survey)
quasars, contrary to results found for radio-loud AGN.  Studies of
optical emission lines in quasars have revealed a striking set of
correlations between various emission line properties, known as the
Boroson \& Green eigenvector~1. Until recently it was generally
accepted that eigenvector~1 does not depend on orientation as it
strongly correlates with [OIII]\,$\lambda$5007 emission, thought to be
an isotropic property. However, recent studies of radio-loud AGN have
questioned the isotropy of [OIII] emission and concluded that
[OII]\,$\lambda3727$ emission is isotropic. In this paper we
investigate the relation between eigenvector~1 and [OII] emission in
radio-quiet BQS quasars, and readdress the issue of orientation as the
driver of eigenvector~1. We account for the small blue bump present at
[OII] wavelengths and subtract FeII emission that contaminates [OIII]
emission. We find significant correlations between eigenvector~1 and
orientation independent [OII] emission, which implies that orientation
does not drive eigenvector~1. The luminosities and equivalent widths
of [OIII] and [OII] correlate with one another, and the range in
luminosities and equivalent widths is similar. This suggests that our
radio-quiet quasar sample of the Bright Quasar Survey is largely free
of orientation dependent obscuration and/or ionization effects.  We
conclude that neither the [OIII] emission nor the [OII]/[OIII] ratio
are dependent on orientation in radio-quiet quasars, contrary to
recent results found for radio-loud quasars.

\end{abstract}
 

\section{Introduction}

Studies of optical emission lines in quasars have revealed some
striking correlations that may well be related to the fundamental
properties of the accreting black hole system. Boroson \& Green (1992,
hereafter BG92) performed a principal component analysis (PCA) on the
BQS quasar sample (Schmidt \& Green 1983) and showed
that the primary eigenvector (hereafter eigenvector~1 or EV1), which
was responsible for $\sim$ 30\% of the variance in the data, was
anticorrelated with various measures of FeII\,$\lambda 4570$ strength
(equivalent width and FeII/H$\beta$ ratio), correlated with
[OIII]\,$\lambda 5007$ strength (luminosity and peak) and H$\beta$
FWHM, and anticorrelated with the blue asymmetry of the H$\beta$
line. It was later found that these optical line properties correlate
with UV properties: CIII] width, SiIII]/CIII] ratio, CIV and NV
strength (Wills et al. 1999; Kuraszkiewicz et al. 2000) and with soft
X-ray properties: luminosity and spectral index (Boroson \& Green
1992; Corbin 1993; Laor et al. 1994, 1997). Recently Brandt \& Boller
(1998) showed that the correlations between EV1 and the X-ray
properties are stronger than those with the individual line
parameters, suggesting that the EV1 has a more fundamental physical
meaning. A number of physical parameters have been suggested to drive
EV1 including accretion rate, orientation, and black hole spin.

BG92 and Boroson (1992) argued that EV1 is not driven by an
orientation effect (i.e. some anisotropic property), despite the
strong dependence on H$\beta$ line width, as it is strongly correlated
with the [OIII]\,$\lambda 5007$ (hereafter [OIII]) luminosity, which
was assumed to be isotropic.  However, the isotropy of the [OIII]
emission in other AGN has since been called into question. Jackson \&
Browne (1990) studied a sample of powerful narrow-line radio galaxies
and radio-loud quasars, which, in the context of Unified Models
(e.g. Antonucci 1993), are considered to be the same type of object
viewed from different angles to the radio axis. The [OIII] line
luminosity of the narrow-line radio galaxies (viewed edge-on) is lower
by 5--10 times than that of the quasars, matched in redshift and
extended radio luminosity. This result was surprising. It was expected
that the [OIII] emission would be the same in both samples, as it was
thought to originate from distances large enough to be unaffected by
obscuring material from the dusty torus.  Hes, Barthel \& Fosbury
(1996) found that radio-loud quasars and powerful narrow-line radio
galaxies show no difference in [OII]\,$\lambda3727$ (hereafter [OII])
emission, suggesting that [OII] emission, and not [OIII] emission, 
is isotropic. As [OIII] has a higher critical density and higher
ionization potential, and hence lies nearer to the central engine,
this difference can be explained if the [OIII] emission region extends
to sufficiently small radii to be obscured by the dusty torus when the
active nucleus is viewed ``edge-on''. Support for this scenario was
provided by the detection of [OIII] emission in polarized light in 4
out of 7 radio galaxies (one also showing [OII] polarization) while a
sample of radio-loud quasars showed none (Di Serego Alighieri et
al. 1997). Polarized [OIII] emission has also been observed in
NGC~4258 (Wilkes et al. 1995; Barth et al. 1999) a Seyfert galaxy with
an edge-on molecular disk surrounding the nucleus. Baker (1997),
studying a complete sample of low frequency radio selected quasars
from the Molonglo Quasar Sample, found that the [OII] to [OIII] ratio
is anticorrelated with the radio-core to lobe flux density ratio $R$,
which is generally used as an orientation indicator. This again
implies that [OIII] is affected by dust absorption as the orientation
becomes more edge-on. Similarly, the [OIII] luminosity versus radio
luminosity correlation shows a larger scatter than the similar [OII]
versus radio correlation (Tadhunter et al. 1998) in the 2 Jy extended
radio selected sample (Wall \& Peacock 1985). This additional scatter
could again be explained by dust obscuration of the [OIII] emission
although the authors prefer an interpretation in terms of the higher
sensitivity of [OIII] to the ionization parameter (Tadhunter et
al. 1998).

If the central regions of radio-loud quasars and powerful radio
galaxies are basically similar to the central regions of radio-quiet
quasars (with the exception of the existence of the radio jets) then
by analogy we would expect the behavior of the [OII] and [OIII] lines
to be similar in both classes. Indeed, Seyfert~1 galaxies, which in
the Unified Model scenario correspond to the face-on Seyfert~2
galaxies, have higher [OIII] luminosities than Seyfert~2s with
comparable radio luminosity (Lawrence 1987; but see Keel et al. 1994
who find no difference in a sample of IRAS selected Seyfert
galaxies). This suggests that L([OIII])/L([OII]) could be an
orientation indicator not only in radio-loud but also in radio-quiet
quasars. Similarly FeII emission strength and the broad line widths
are strongly dependent on orientation in radio-loud QSOs (e.g.  Miley
\& Miller 1979; Wills \& Browne 1986; Vestergaard, Wilkes \& Barthel
2000) with stronger FeII and narrower lines in face-on sources. Again,
by analogy, this suggests that the extreme EV1 objects, Narrow-Line
Seyfert 1s (NLS1), which have stronger FeII emission and narrow lines,
could also be face-on.

Given the strong evidence for anisotropic [OIII] emission in
radio-loud quasars, we present an investigation of the behavior of
[OII] emission in a radio-quiet subset of the optically-selected
Palomar BQS sample to study the [OII] relation to EV1 in comparison with
that of [OIII]. This allows us to revisit the question of orientation
as a driver of EV1 and will lead to a better understanding of the
underlying physics driving the strongest set of emission line
correlations found to date for quasars, and so provide information on
their central regions. We also compare the [OII] and [OIII] emission
in our radio-quiet, optically selected sample with radio-loud samples
to investigate whether the behavior of [OII] and [OIII] emission is 
similar in the two classes. Finally, we address the question of whether
the [OII]/[OIII] ratio is an orientation indicator in radio-quiet
quasars.

\section{Observations and Data Reduction}   

\begin{figure} [t!]
\vspace{9.0truecm}
{\includegraphics{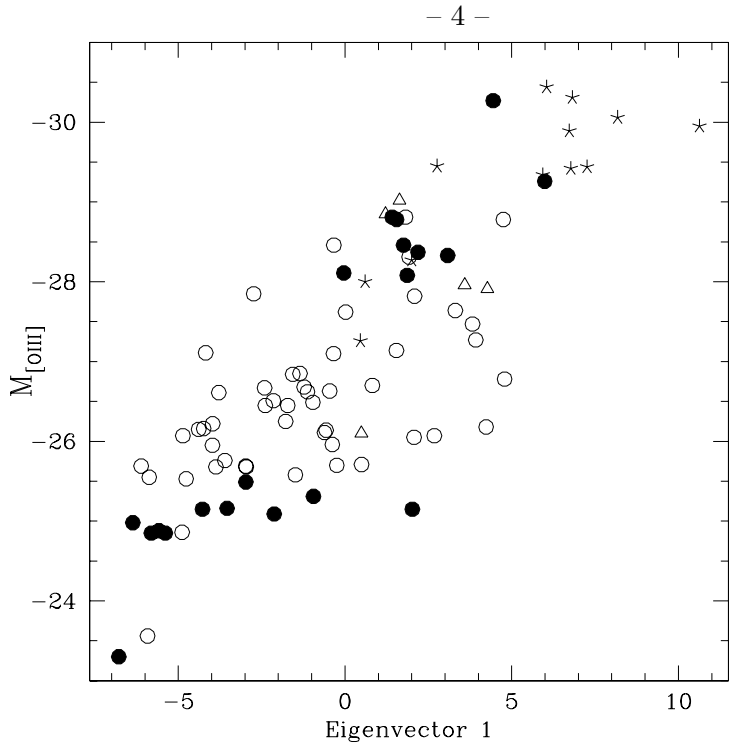}}
\caption{The [OIII] absolute magnitude ($M_{\rm [OIII]}$) from BG92
versus BG92 eigenvector~1. Filled circles indicate the radio-quiet
quasars in the current sample, open circles: radio-quiet QSOs in BG92,
stars: steep spectrum radio-loud QSOs, and open triangles: flat
spectrum radio-loud QSOs.
\label{fig:sample}} 
\end{figure}

In order to carry out this investigation the sample needs to cover a
wide range of EV1 values. Since the [OIII] luminosity ($M_{\rm
[OIII]}$) is directly measurable and strongly correlates with EV1, our
objects were selected to have either high or low [OIII] luminosity
i.e.  $M_{\rm [OIII]} < -28$ or $M_{\rm [OIII]} > -25.5$ in the BG92
sample (see Figure~\ref{fig:sample}; however a few radio-quiet quasars
satisfying these selection criteria were not included in our sample
due to lack of observing time). Our sample consists of 11 objects with
low [OIII] luminosity and 9 objects with high [OIII] luminosity. We
have obtained high signal-to-noise spectra of these quasars which
include both the [OIII]\,$\lambda$5007\AA\ and
[OII]\,$\lambda$3727\AA\ lines. The observations were made between
1997 June and December with the FAST spectrograph on the 1.5~m
Tillinghast telescope on Mt. Hopkins in Arizona. A 300 gpm grating set
to cover the wavelength range 3500$-$7500\AA\ was used with a
$2^{\prime\prime}$ aperture, yielding a resolution of $\sim 6$\AA.
Spectrophotometry was carried out, in photometric conditions, by
observing each quasar twice, first through a large $5^{\prime\prime}$
aperture and second through the small, $2^{\prime\prime}$, aperture
with a longer exposure time to obtain high spectral resolution and
signal-to-noise. A standard star was observed through the wide
aperture, at similar air mass, immediately before or after the quasar
observation, to provide flux calibration. The data were reduced in the
standard manner using IRAF\footnote{IRAF (Image Reduction and Analysis
Facility) is distributed by the National Optical Astronomy
Observatories, which are operated by AURA, Inc., under cooperative
agreement with the National Science Foundation.} (see Tokarz \& Roll
1997 for details).  The continuum of the small aperture data was then
normalized to match the shape and absolute flux level of the large
aperture observation yielding a final spectrum with a (judged from our
experience) photometric accuracy of $\sim 5$\%.  The observational
details are given in Table~1 and the calibrated spectra are presented
in Figure~\ref{fig:spectra}.

\begin{figure} [t!]
\vspace{9.0truecm}
{\includegraphics{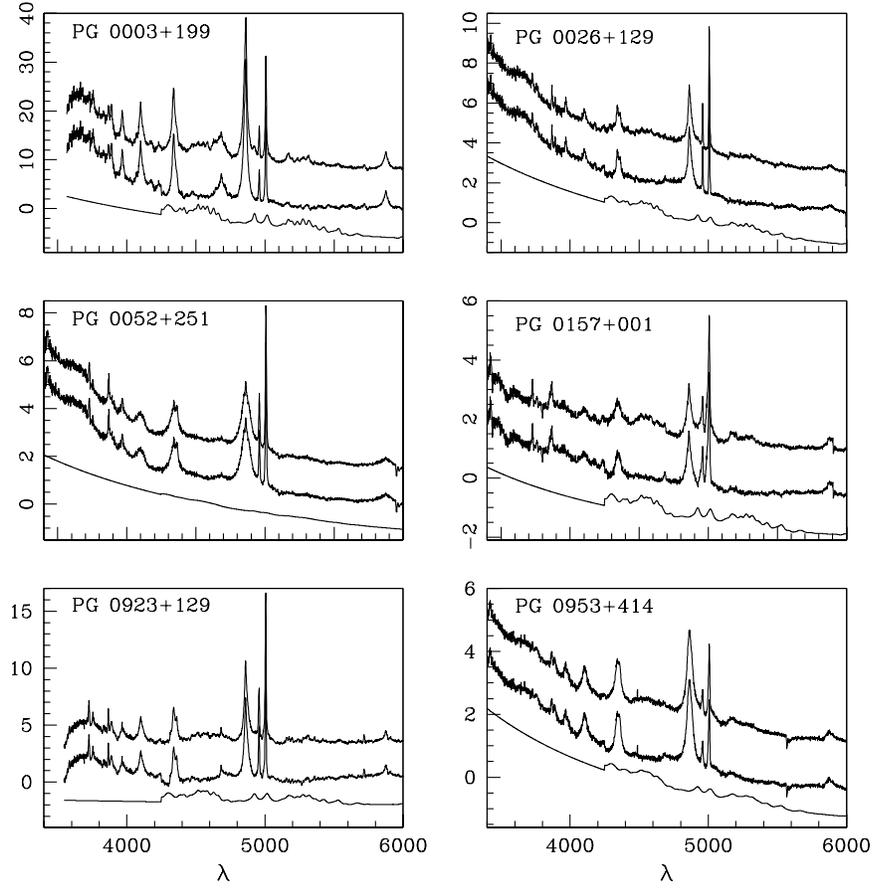}}
\figcaption{The observed and FeII-subtracted spectra in the
rest frame, in order of increasing right ascension. Only the
wavelength range around the [OII] and [OIII] lines is presented on a
scale $F_{\lambda}$ (in units of $10^{-15}$ erg s$^{-1}$ cm$^{-2}$ 
\AA$^{-1}$) as a function of $\lambda$ (\AA). The top spectrum is the
observed spectrum, middle is the FeII subtracted spectrum and the
bottom spectrum is the fitted FeII model. The FeII subtracted  spectrum
and the FeII model have been shifted downwards by an arbitrary value
for clarity.
\label{fig:spectra}}
\end{figure}


\begin{figure} [t!]
\vspace{9.0truecm}
{\includegraphics{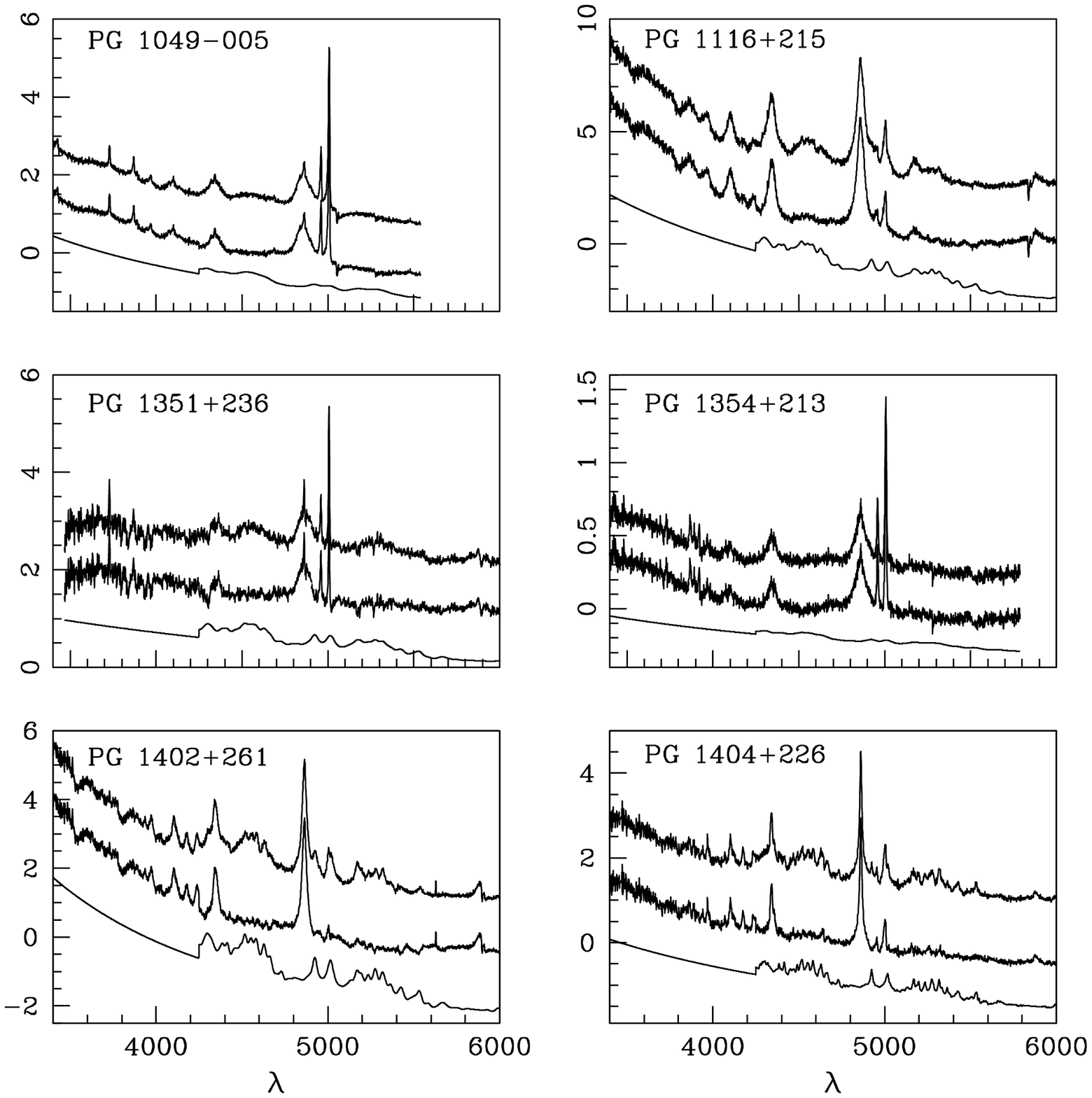}}
\addtocounter{figure}{-1}
\figcaption{\it $-$ continued}
\end{figure}

\clearpage

\begin{figure} [t!]
\vspace{9.0truecm}
{\includegraphics{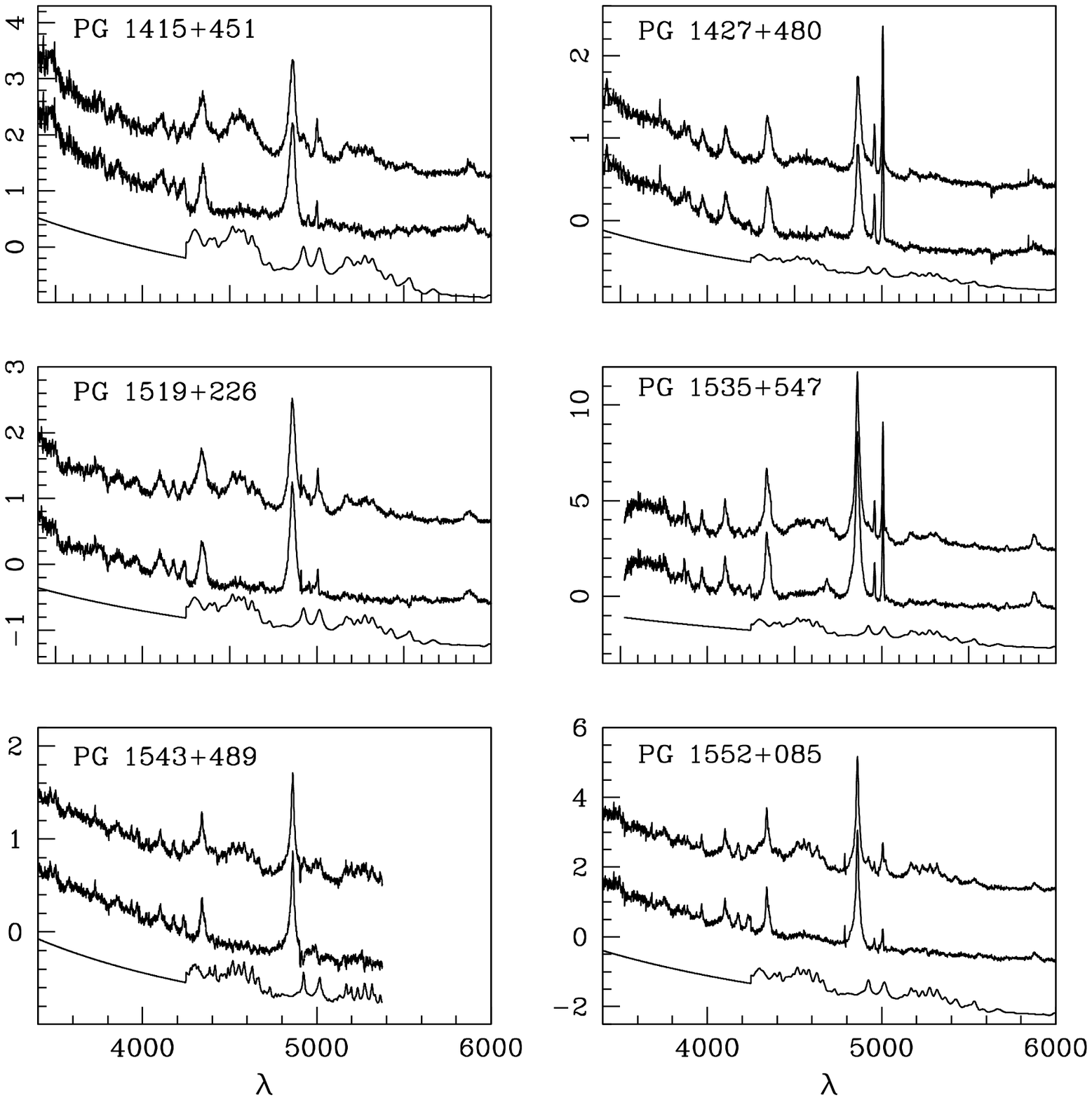}}
\addtocounter{figure}{-1}
\figcaption{\it $-$ continued}
\end{figure}

\clearpage

\begin{figure} [t!]
\vspace{9.0truecm}
{\includegraphics{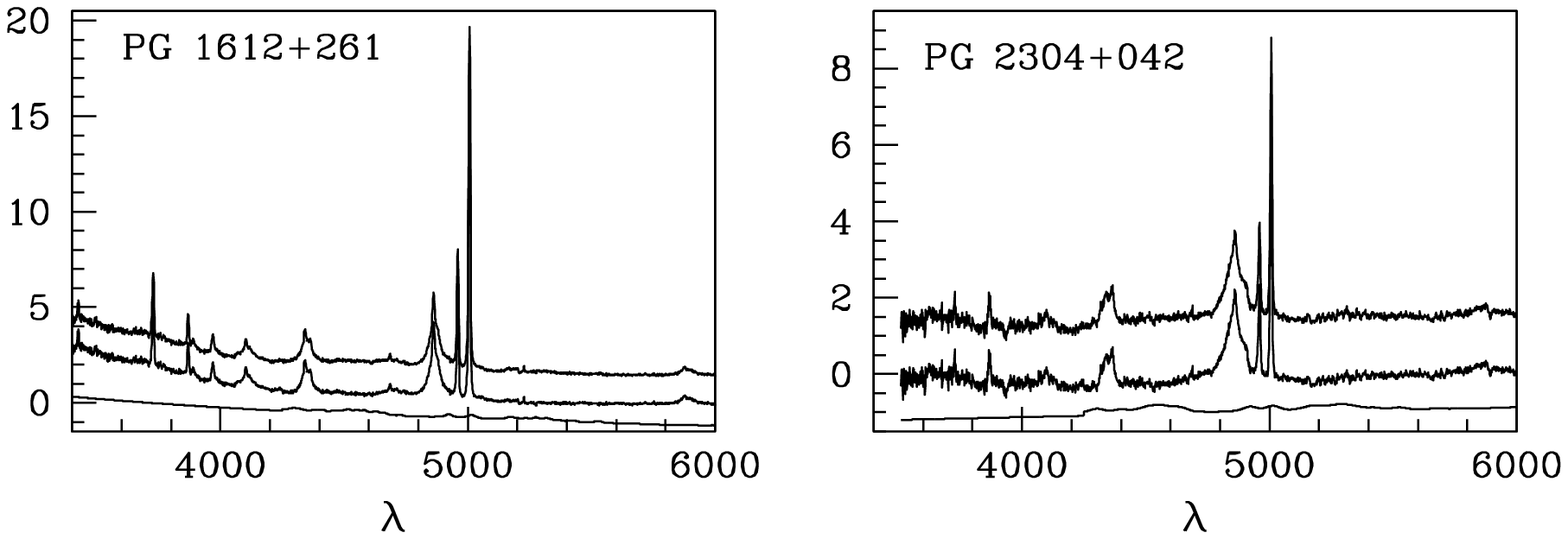}}
\addtocounter{figure}{-1}
\figcaption{\it $-$ continued}
\end{figure}

\subsection{FeII subtraction}

It has been shown (e.g. BG92) that FeII emission is present in the
form of broad humps of blended lines at $\lambda\lambda$4450$-$4700 and
$\lambda\lambda$5150$-$5350. This emission severely complicates
measurements of emission line strengths and widths. Around the
[OIII]\,$\lambda\lambda$4949,5007 lines, a strong FeII optical
multiplet (42) consisting of 3 lines at
$\lambda\lambda$4924,5018,5169 contaminates the [OIII]
emission. This contamination is particularly strong in the weak [OIII]
sources since, following EV1, these objects usually have 
strong FeII emission (see BG92). At the [OII]\,$\lambda$3727
wavelength a ``small bump'' (spanning 2000$-$4000\AA, which is a
blend of FeII lines with the Balmer continuum emission) resides, which
has to be taken into account when making measurements of the [OII]
line.

In our spectra we have fitted the underlying continuum with a
power-law using only those regions of the spectrum uncontaminated by
the FeII emission ($\lambda\lambda$4150$-$4270 and
$\lambda\lambda$6160-6280). The underlying power-law continuum was
then subtracted from each spectrum and the FeII emission was modeled
using an optical ($\lambda\lambda4247-7000$) FeII template, kindly
provided by T. Boroson (BG92). The template FeII spectrum was
broadened by convolving with Gaussian functions to multiple widths
starting at 1000~km~s$^{-1}$ and separated by steps of
250~km~s$^{-1}$. A two-dimensional iron emission model was constructed
with line width as one dimension and rest wavelength as the other (see
Vestergaard \& Wilkes 2000). The iron emission in the object's spectrum
was fitted by scaling this 2-D iron model to the iron emission on
either side of the H$\beta$ and [OIII] lines.
%
%
As a check-up and confirmation
of this primary normalization, five additional scalings from 0.6 to
1.4 in steps of 0.2 were applied to each normalized iron model and
these were also compared to the object's spectrum. These additionally
scaled models were never needed, confirming that the primary
normalization was satisfactory and adequate in each case. $\chi ^2$
statistics and residual flux measurements were used to determine the
best fit iron model, but manual inspection of how each model fits the
spectrum and of the residual spectra was also necessary and was of
high importance due to the complexity of QSO spectra. Once the
best-fitting FeII model was identified it was subtracted from the
original QSO spectrum allowing for an improved underlying continuum to
be determined. The iteration over the FeII model and continuum setting
(Vestergaard \& Wilkes 2000) was continued until little change was
seen from one step to next. Usually no more than 2 iterations were
needed. 

The best-fitting FeII emission model was then subtracted from the QSO
spectrum. The previously subtracted power-law continuum was then added
back into the spectrum giving an FeII-subtracted spectrum. We present
these spectra in Figure~\ref{fig:spectra} along with the fitted FeII
models to each spectrum and the original, uncorrected QSO spectra.

\subsection{Line Measurements}

The fluxes and equivalent widths (EW) of the [OIII]\,$\lambda5007$, 
[OII]\,$\lambda3727$, H$\beta$\,$\lambda$4861 and FeII\,$\lambda4570$
lines (for comparison with the BG92 data) in the FeII subtracted
spectra were measured using the {\it splot} task in IRAF. For [OIII]
and H$\beta$ we took the previously fitted, underlying, power-law
continuum and integrated the spectrum above this continuum and across
the observed emission line (we used keystroke `e' in the IRAF {\it
splot} task). The flux and equivalent width of the FeII\,$\lambda4570$
optical multiplet were measured in the $\lambda\lambda$4434$-$4684
range across the fitted FeII emission models. The equivalent widths and
line luminosities of the emission lines are presented in Table~2. Our
measurements of equivalent widths of H$\beta$, [OIII], FeII lines and
FeII/H$\beta$ agree with BG92 to within 30\%, except where noted 
in the table.

\begin{figure} [t!]
\vspace{9.0truecm}
{\includegraphics{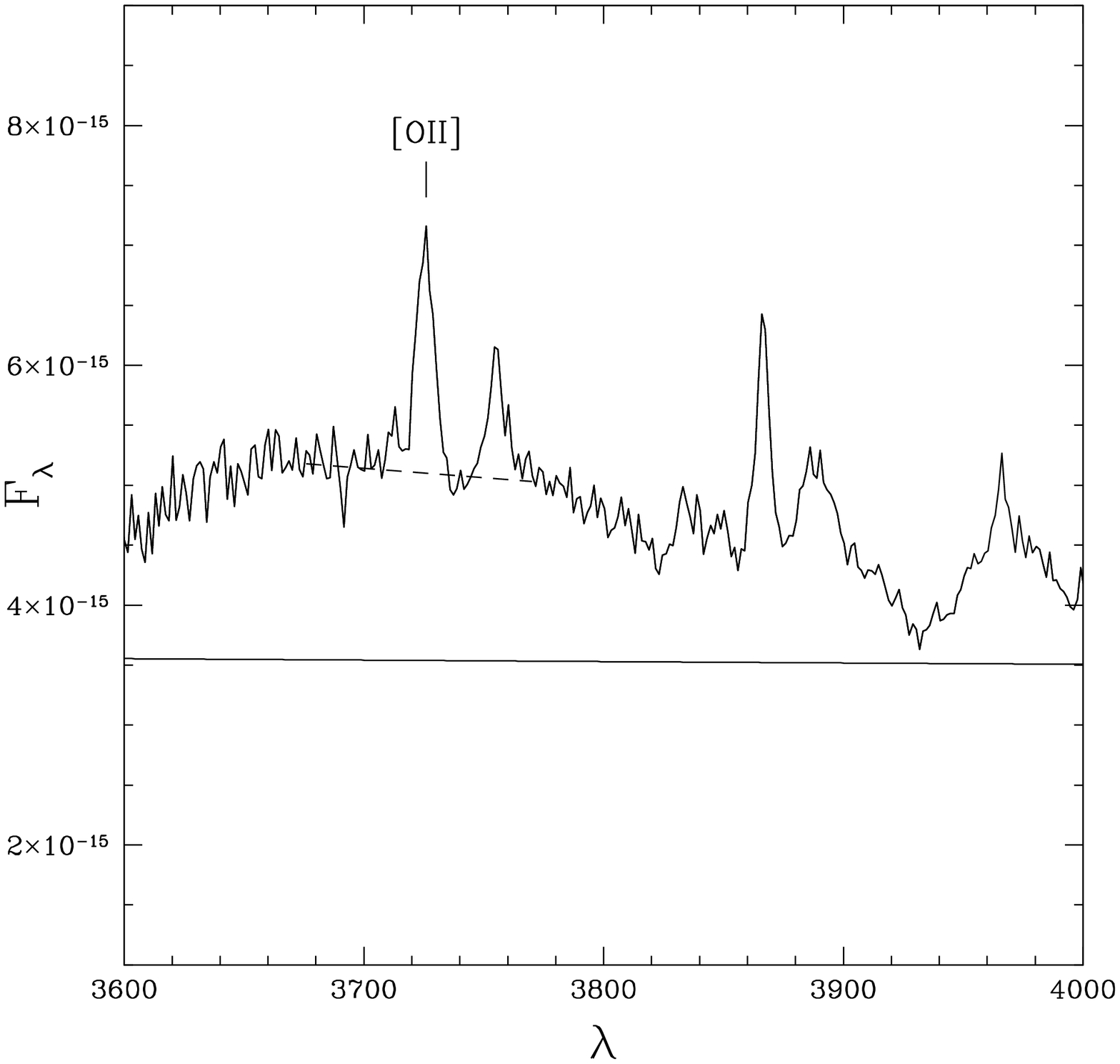}}
\figcaption{The spectrum of PG~0923+129 around the [OII]
line. The local continuum representing the continuum from
the ``small bump'' is indicated by a dashed line, while the power-law
continuum fitted to the whole spectrum and lying well below the 
``small bump'' is shown by a solid line.
\label{fig:small_bump}}
\end{figure}

\begin{figure} [t!]
\vspace{9.0truecm}
{\includegraphics{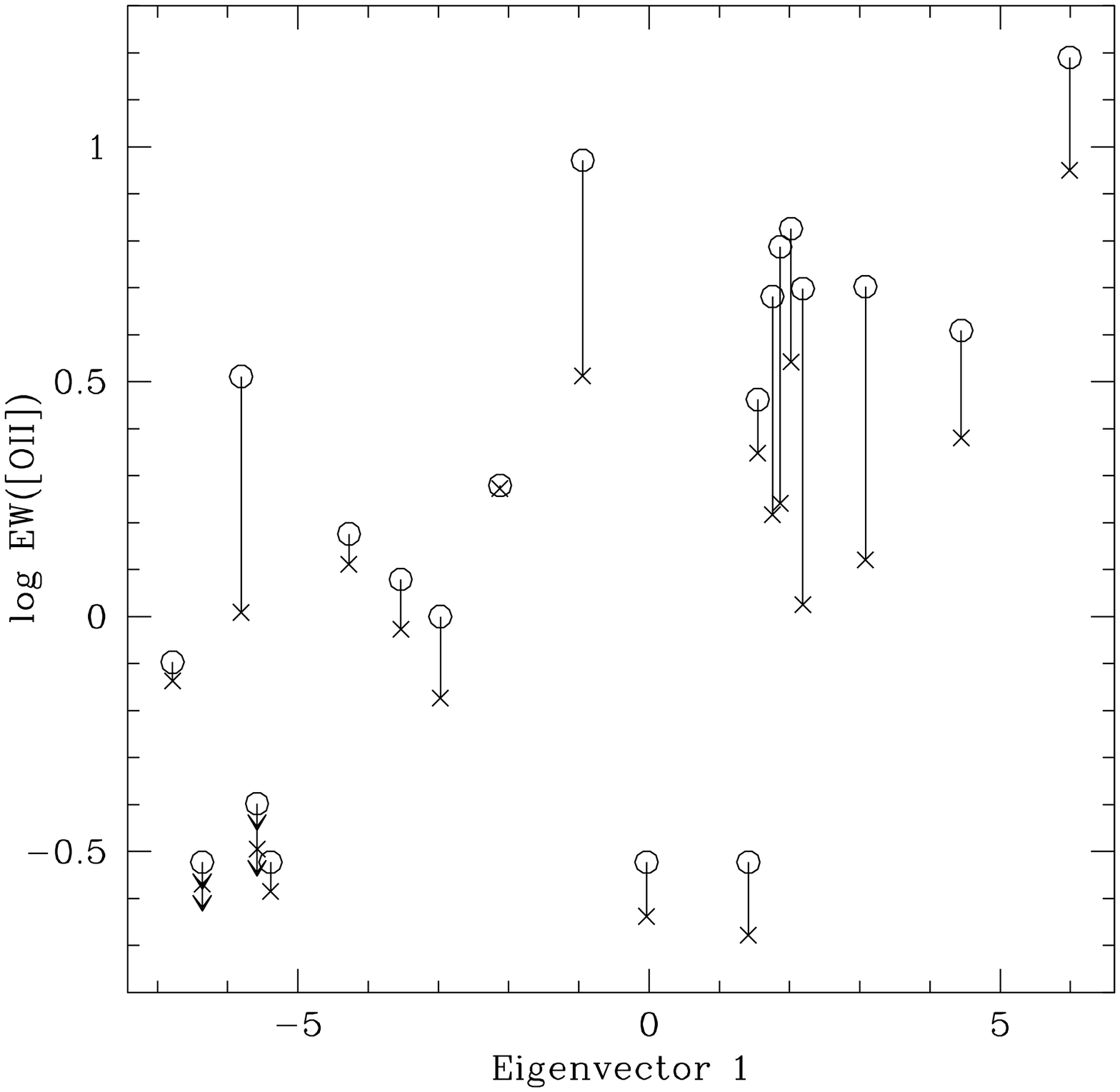}}
\figcaption{The [OII] equivalent width (EW) versus the Boroson \& Green
(1992) eigenvector~1. Circles denote our equivalent width measurements
with respect to our fitted underlying continuum, crosses denote
equivalent width measurements relative to the local continuum
i.e. with the small bump not taken into account.
\label{fig:EW_OII}}
\end{figure}

The flux of the [OII]\,$\lambda3727$ line was measured by integrating
the spectrum above a ``local'' continuum i.e. the ``small bump''
(dashed line in Figure~\ref{fig:small_bump}; we did not subtract the
``small bump'' as it is not included in the template). The equivalent
width was then defined as the ratio of the flux of the [OII] line
estimated above the ``small bump'' to the flux over the same
wavelength range in the power-law continuum (solid line in
Figure~\ref{fig:small_bump}; this took care of the contamination from
the ``small bump'').  If instead the [OII] flux was divided by the
local underlying continuum, as is more usual, then the equivalent
width would be underestimated (by factors up to $\sim$ 5). A
comparison of the equivalent widths obtained by the two methods
(Figure~\ref{fig:EW_OII}) illustrates a significant systematic shift
in equivalent widths whose magnitude varies from source to source,
emphasizing the need to take into account the small bump when
measuring the [OII] line.

Our sample consists of luminous quasars ($M_B < -23$), which are
generally not highly variable at optical wavelengths (e.g. Giveon et
al. 1999).  However, if any of the quasars had undergone a change of
continuum from the time of BG92 observations, it would be seen in the
differing EW([OIII]) measurements (as the [OIII] emitting region lies
far enough from the central engine to be unaffected by the changing
continuum on time scales of a few years). This may be the case for the
following objects: PG~0026+129, PG~1427+480, where our EW([OIII])
measurements are lower than those of BG92 and in PG~1354+213, where
our EW([OIII]) is higher (see Table~2). These objects also show lower
(or higher respectively) EW(H$\beta$) and slightly lower/higher
EW(FeII) measurements (although the change is less than 30\% and hence
not noted by `*' in Table~2). The FeII/H$\beta$ ratio did not differ
significantly from BG92 in these objects.  For PG~1543+489 BG92 quote
EW([OIII])=0, while we were able to measure this line in our spectra
and obtained a value of 4.8\AA.  A comparison of our line equivalent
width measurements in PG~1612+261 and PG~2304+042 indicates that we
have set the underlying continuum higher for PG~1612+261 and lower for
PG~2304+042, and with a flatter slope, probably due to the wider
wavelength range covered by our spectra, allowing a better continuum
determination.

The line equivalent width measurements are influenced by the choice of
aperture. BG92 used $1.^{\prime\prime}5$ aperture, while we use
spectra obtained with a $2^{\prime\prime}$ aperture and flux
calibrated using quasar spectra through 
a $5^{\prime\prime}$ aperture (which in turn reference a star
through a $5^{\prime\prime}$ aperture). The amount of
starlight for many PG quasars has been measured by McLeod \& Rieke
(1994a,b). We found that for most of the objects in our sample the
starlight contribution is of the order of 20\% of the total flux in
the H (1.65$\mu$m) band i.e. 13\% at 4000\AA, assuming a starlight
template from Elvis et al. (1994). However, PG~0157+001 has a 43\%
starlight contribution at band H (i.e. 29\% at 4000\AA) and is also
spatially extended ($12^{\prime\prime} \times
12^{\prime\prime}$). Assuming a uniform distribution of starlight, and
a constant AGN energy output, we can roughly estimate the starlight
contribution in this (worst case scenario) object, which is 7\% ($0.43
\times 5^2/12^2$) at H band and 5\% ($0.29 \times 5^2/12^2$) at
4000\AA\ in our spectra, and 0.7\% and 0.5\% respectively in BG92
spectra.  Hence the level of starlight contamination in ours and the
BG92's equivalent widths is well below the typical 30\% errors 
due to other factors such as continuum placement and line measurement.

\subsection{Eigenvector~1}

\begin{figure} [t!]
\vspace{9.0truecm}
{\includegraphics{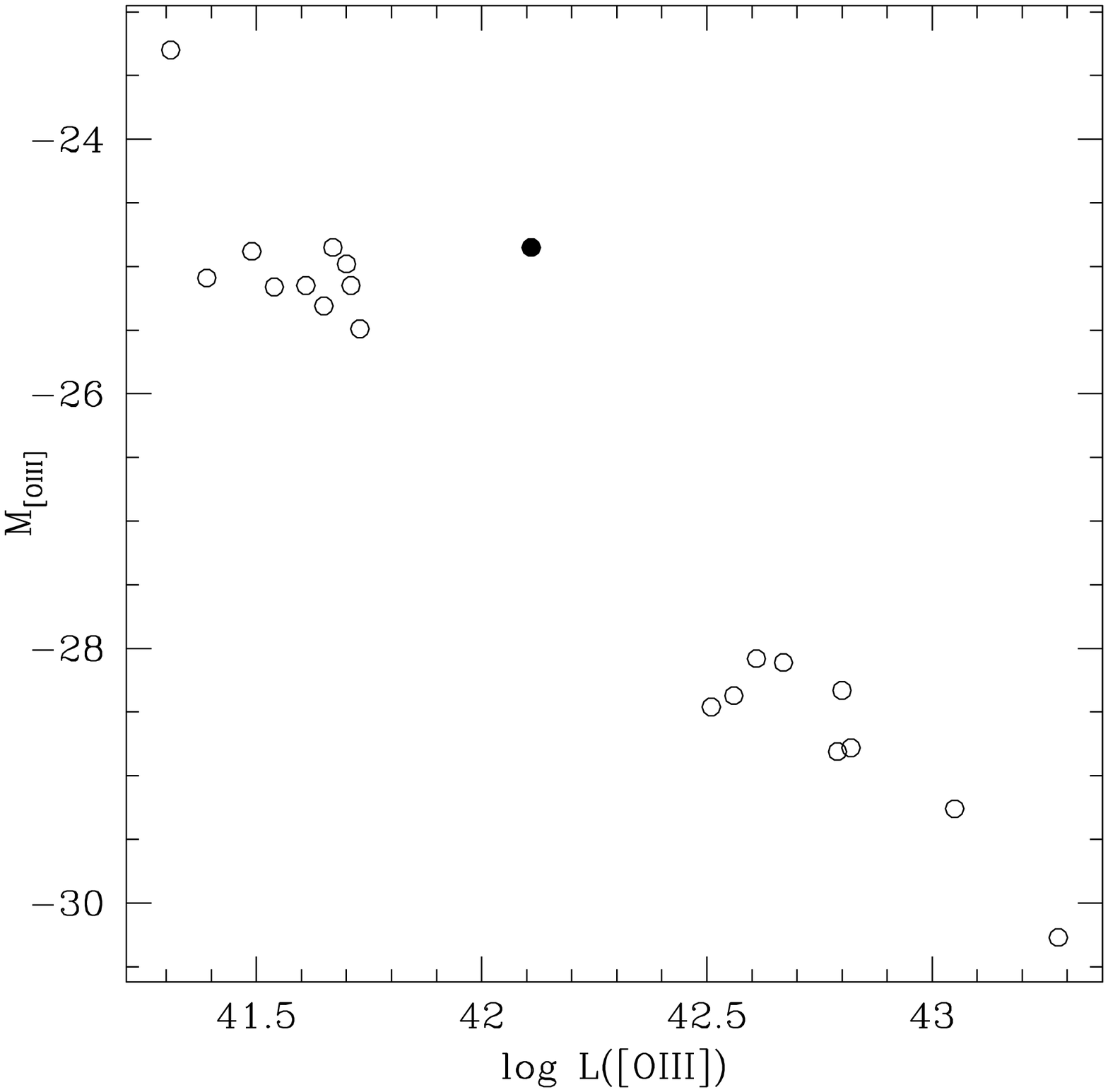}}
\figcaption{The [OIII] absolute magnitude ($M_{\rm [OIII]}$) from Boroson
\& Green (1992) versus our [OIII] luminosity measurements
demonstrating good agreement. The filled circle represents
PG~1543+489, for which the Boroson \& Green measurement of EW([OIII])
is inconsistent with ours.
\label{fig:MOIII_LOIII}}
\end{figure}

\begin{figure} [t!]
\vspace{9.0truecm}
{\includegraphics{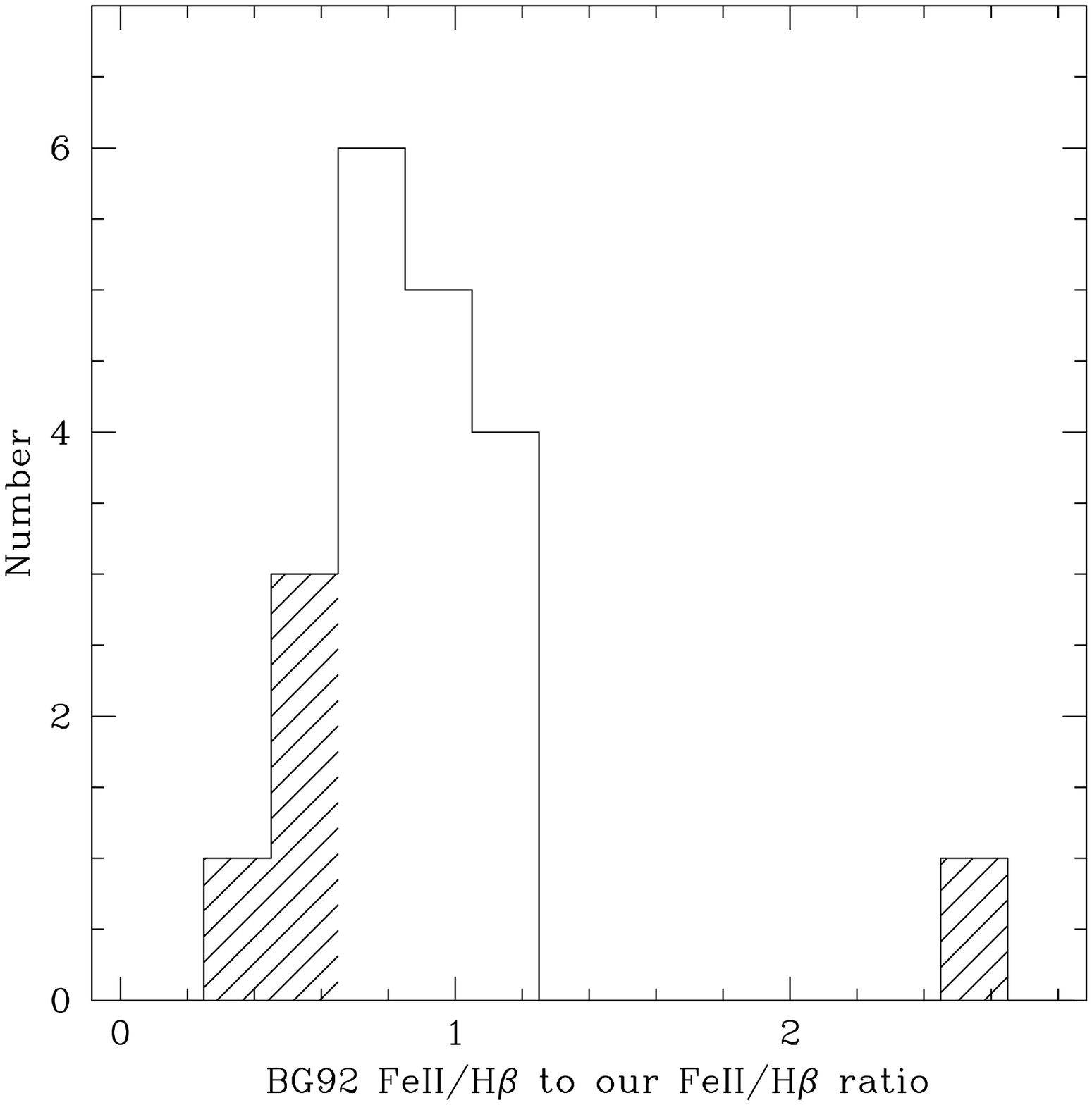}}
\figcaption{A histogram of the ratio of FeII/H$\beta$ measured by Boroson \&
Green (1992) to the FeII/H$\beta$ ratio measured by us. Shaded areas 
denote quasars which FeII/H$\beta$ measurements differed from 
Boroson \& Green (1992) by at least 30\%.
\label{fig:FeIIour_BG}}
\end{figure}

The EV1 values for our sample QSOs (kindly provided by T. Boroson)
were calculated by applying the PCA analysis to the BQS QSOs sample.
We quote these values (after BG92) in the last column of
Table~2.  EV1 was shown by BG92 to depend strongly on the peak and
absolute magnitude of the [OIII] line ($M_{\rm [OIII]}$) and the
FeII/H$\beta$ ratio. 
In general, our line measurements agree well (to within 30\%) with
those of BG92, implying that it is valid to use the EV1 values from
BG92.  In Figure~\ref{fig:MOIII_LOIII} we present a comparison of the
[OIII] luminosity measured by us and the absolute magnitude $M_{\rm
[OIII]}$ from BG92, defined as $M_{\rm V}-2.5\log EW([OIII])$. The
general agreement is good with only one highly discrepant object:
PG~1543+489 (indicated in Figure~\ref{fig:MOIII_LOIII} by a filled
circle; see Section~2.2 for further explanation of the differences
between our and BG92 measurements). In four other objects
(PG~0157+001, PG~0953+414, PG~1612+261 and PG~2304+042) we measured
the FeII/H$\beta$ ratio to be significantly larger ($ > 30\%$) than in
BG92 (see Table~2 and Figure~\ref{fig:FeIIour_BG}).  A comparison of
our line equivalent width measurements in these objects indicate that
we have set the underlying continuum lower and with a flatter slope in
our objects, probably due to the fact that our spectra cover a larger
wavelength range and that we iterated over the continuum setting and
FeII models in the FeII subtraction process.  In PG~0052+251 the
FeII/H$\beta$ measured by us is smaller than in BG92. However in our
FeII subtracted spectrum we still have some residual
FeII\,$\lambda$4570 emission left (as our primary goal was to optimize
the FeII fit around the [OIII] line and not FeII\,$\lambda$4570), so
the BG92 FeII measurement, and hence the EV1, are probably more
correct.

For the five, discrepant objects, discussed above, we indicate the
direction in which EV1 should move in our figures based on the EV1
range of objects with similar values of FeII/H$\beta$, $M_{\rm
[OIII]}$ and [OIII] peak measurements in BG92. The only way to improve
on this would be to re-run the PCA analysis for the full BG92 sample
using our new values of line measurements, which is beyond the scope
of this paper.

\section{Discussion}

As outlined in the introduction the differences in [OIII] emission
between the narrow-line radio galaxies and radio-loud quasars reported
by Jackson \& Browne (1990) and lack thereof in [OII] (Hes, Barthel \&
Fosbury 1996) suggest that [OIII] is orientation dependent, while
[OII] is more isotropic in the radio-loud AGN. The correlation between
the [OII]/[OIII] ratio and the orientation indicator $R$
reported by Baker (1997), furthermore suggests orientation-dependent
dust obscuration of [OIII] emission and more isotropic [OII]
emission. These results question the BG92 conclusion that EV1 is
independent of orientation based on the assumption of [OIII] isotropy
and allow us to readdress the question of orientation as the driver of
EV1 by studying the dependence of BG92 EV1 on isotropic [OII]
emission. To ensure a wide range in EV1 values we selected
radio-quiet quasars with either high or low [OIII] luminosities
(see Section~2). Under the assumption that our radio-quiet quasars,
which are a subset of the BQS sample, have similar narrow line
emitting regions to the radio-loud quasars and powerful radio-loud
galaxies we presume that [OII] emission is independent of orientation
in radio-quiet quasars. As a result finding a strong relation between
EV1 and [OII] luminosity would imply that EV1 is independent of
orientation (furthermore suggesting isotropic [OIII] emission in
radio-quiet quasars, as [OIII] correlates with EV1) while the lack of
such a relation would suggest that orientation is a factor.

\begin{figure} [t!]
\vspace{9.0truecm}
{\includegraphics{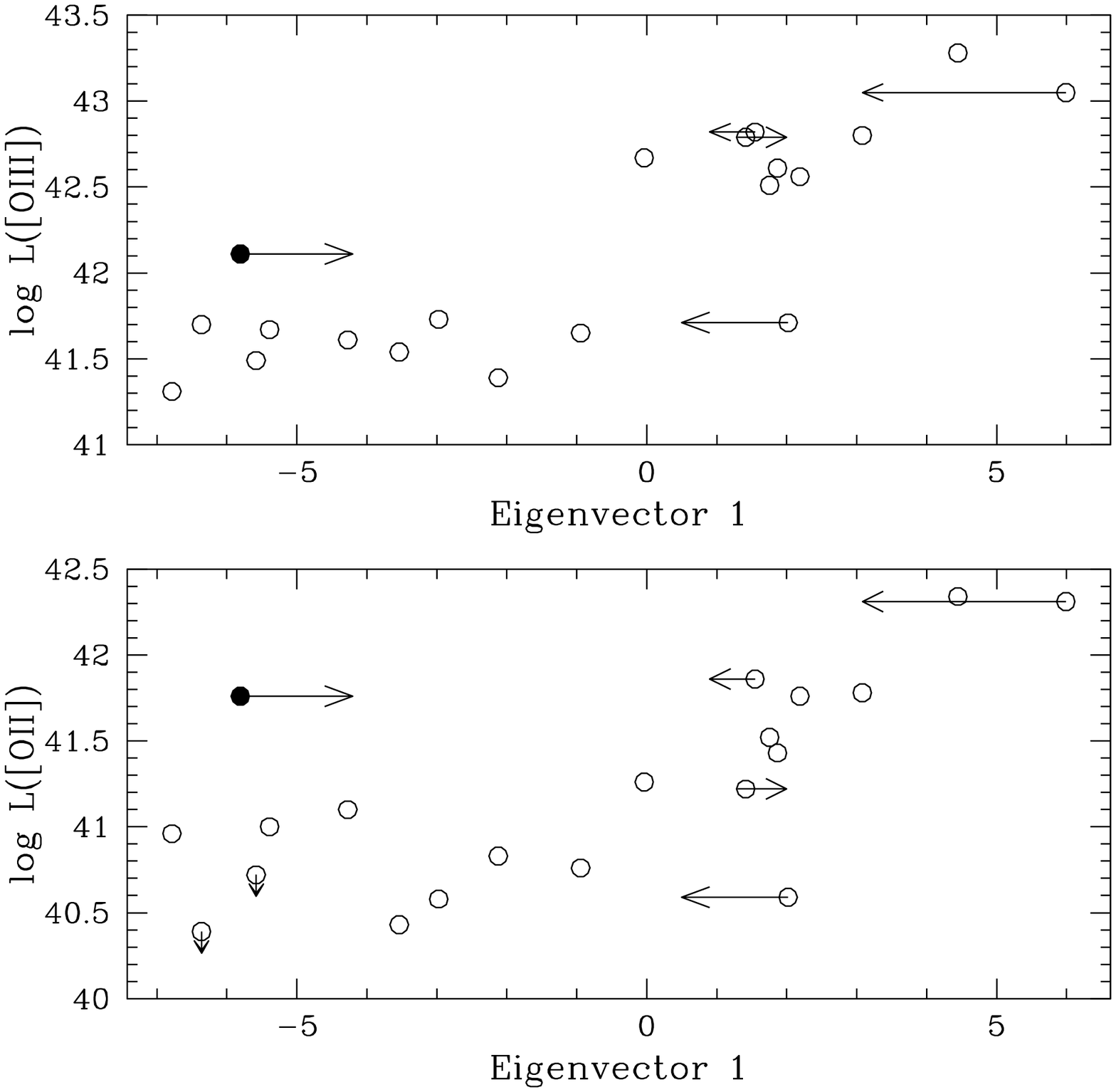}}
\vspace{1.5cm}
\figcaption{The [OIII] and [OII] luminosity versus Boroson \&
Green (1992) eigenvector~1 correlations. PG~1543+489 is shown as a 
filled circle. Arrows indicate a range of EV1 values for PG~0157+001,
PG~0953+414, PG~1543+489, PG~1612+261 and PG~2304+042 based upon our 
differing line measurements in comparison with BG92.
\label{fig:OII_OIII_corr_EV1}}
\end{figure}

\begin{figure} [t!]
\vspace{9.0truecm}
{\includegraphics{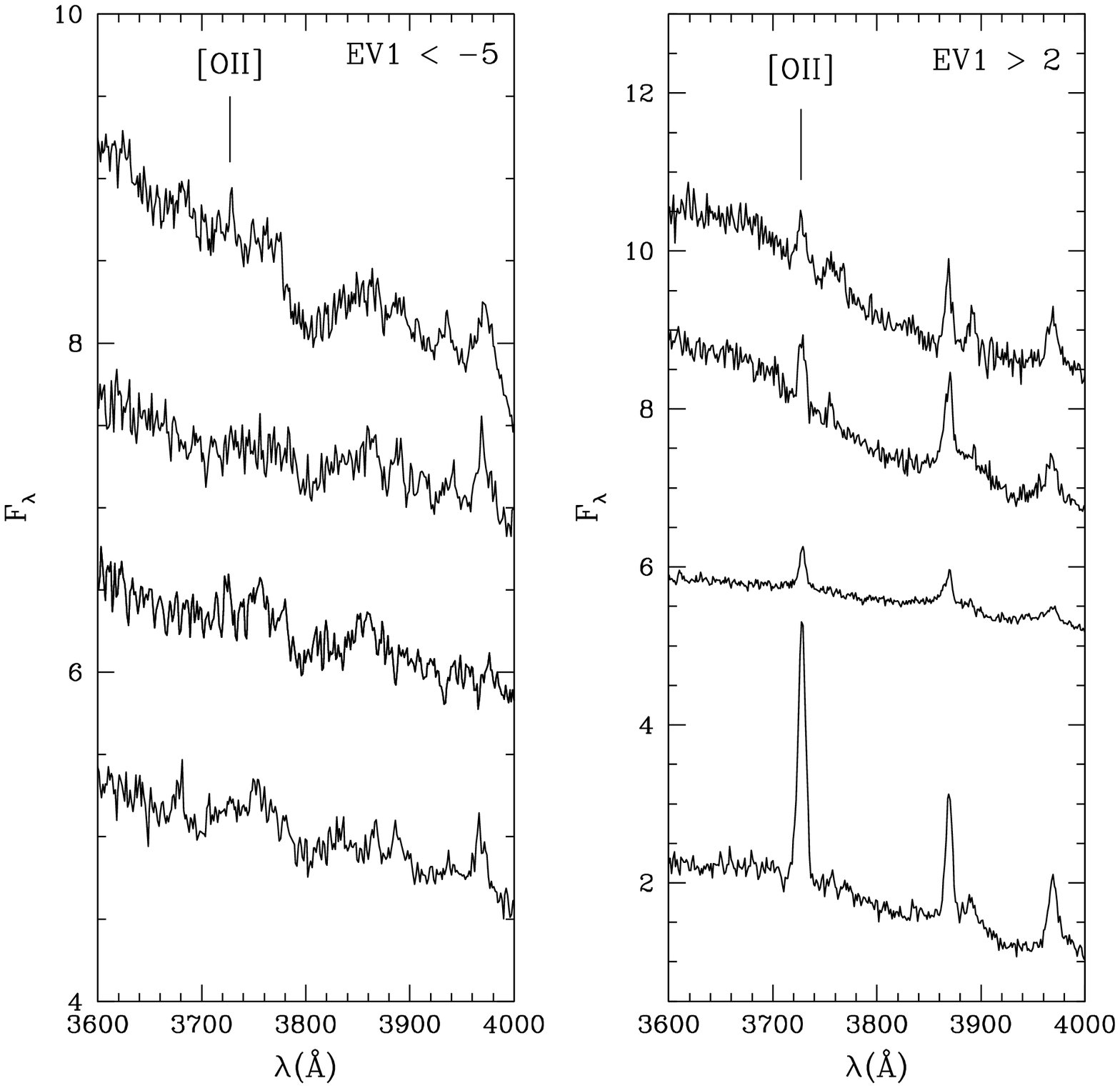}}
\vspace{1.5cm}
\figcaption{The observed spectra in the rest frame showing in detail the
wavelength range around the [OII] line for the most extreme EV1
objects on a scale $F_{\lambda}$ (in units of $10^{-15}$ erg s$^{-1}$ 
cm$^{-2}$ \AA$^{-1}$) as a function of $\lambda$ (\AA). Spectra have
been shifted downwards by an arbitrary value for clarity. The left 
panel shows objects with EV1$< -5$ (from top: PG~1402+261, PG~1404+226, 
PG~1415+451, PG~1552+085) and the right panel objects with EV1
$>2$ (from top: PG~0026+129, PG~0052+251, PG~1049$-$005, PG~1612+261).
\label{fig:spectra_OII}}
\end{figure}

The relations between [OIII] and [OII] luminosities
(L([OIII]), L([OII])) and the Boroson \& Green EV1 are presented in
Figure~\ref{fig:OII_OIII_corr_EV1}.
We find a significant correlation between L([OIII]) and EV1 consistent
with the $M_{\rm [OIII]}$ versus EV1 correlation found by BG92. The
Spearman rank test shows a 0.09\% probability of this correlation
occurring by chance (hereafter we will use $P_S$ to indicate the
chance probability in the Spearman rank test\footnote{We use the ASURV
statistical package (Isobe, Feigelson \& Nelson 1986), which includes
allowance for the presence of upper limits in [OII]
measurements.}). We also find a significant correlation between
L([OII]) and EV1 with $P_S =$ 0.23\%, which becomes stronger with $P_S
=$ 0.08\% if the values of EV1 were updated to allow for the
differences between our measurements and those of BG92 (i.e. values in
the range shown by the arrows in Figure~\ref{fig:OII_OIII_corr_EV1}).
These results imply that EV1 is independent of orientation and suggest
that an intrinsic property, such as the accretion rate onto a black
hole (as suggested by BG92; Pounds, Done \& Osborne 1995; Boller,
Brandt \& Fink 1996; Laor et al. 1997) or the black hole spin (BG92)
may be driving EV1.  In Figure~\ref{fig:spectra_OII} we present the
spectra around the [OII] wavelength for the most positive and the most
negative EV1 objects to show in detail the dependence of [OII] on EV1.

\subsection{Radio-quiet versus Radio-loud Quasars}

The presence of the correlations between L([OII]), L([OIII]) and EV1
found above suggest that in our radio-quiet quasars from the BQS
sample the [OIII] emission is independent of orientation in contrast
to the case of radio-loud quasars.  In order to understand this
apparent dichotomy we study in detail the L([OII]) versus L[OIII]
and EW([OIII]) versus EW([OII]) relations in our radio-quiet sample
and compare it with the radio-loud samples of Baker (1997; hereafter
JB97), and Tadhunter et al. (1998) where orientation combined with
dust or ionization effects (respectively) were found to be present.

\begin{figure} [t!]
\vspace{9.0truecm}
{\includegraphics{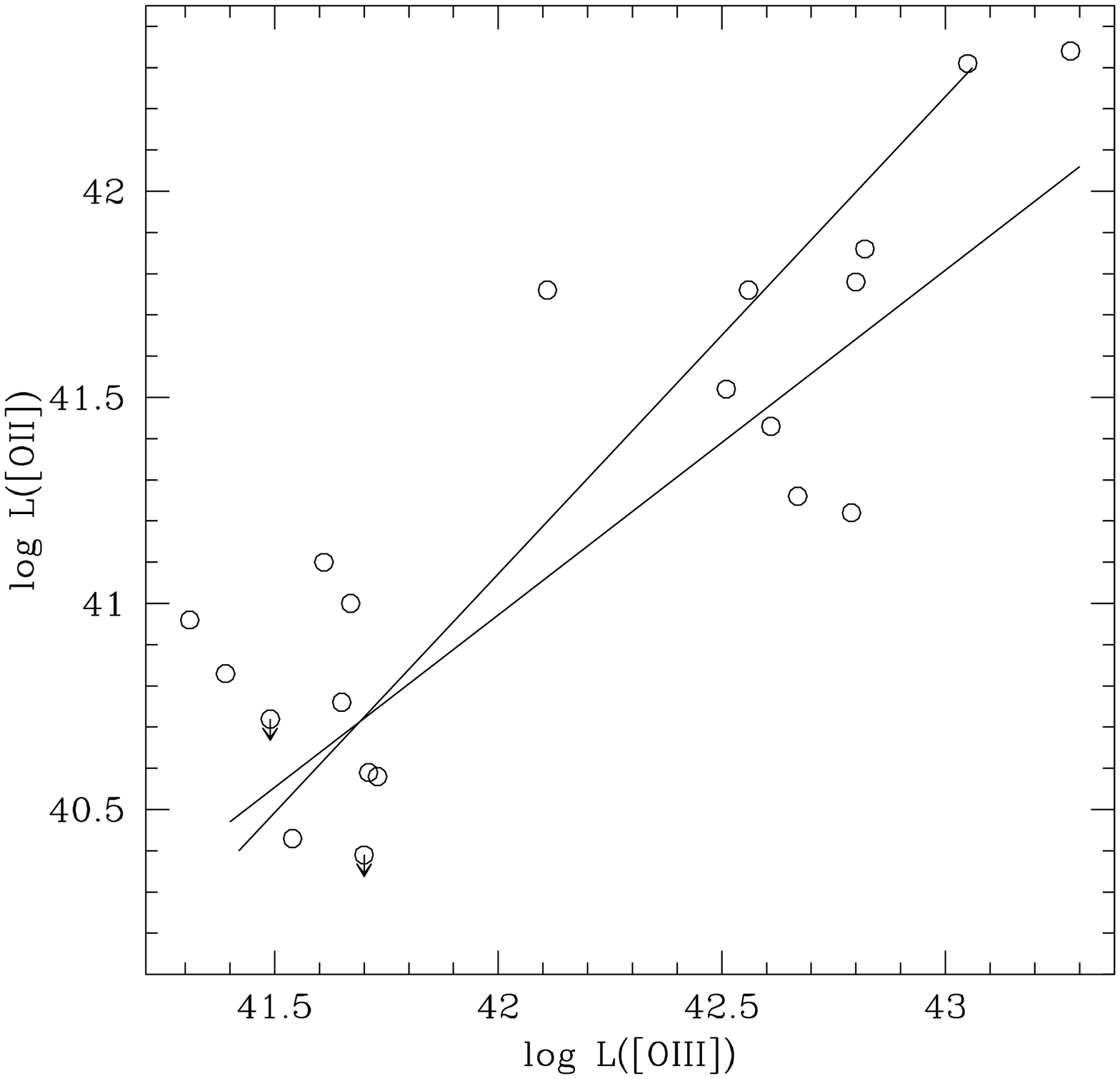}}
\figcaption{The [OII] luminosity versus [OIII] luminosity correlation.
\label{fig:LOII_LOIII}}
\end{figure}

The [OII] and [OIII] luminosities and equivalent widths in our
radio-quiet sample correlate significantly with one another ($P_S$ =
0.04\% for L([OIII]) versus L([OII]) correlation and $P_S$ = 0.31\% for
EW([OIII]) versus  EW([OII]) correlation). Additionally the range in
L([OII]) and L([OIII]) is similar ($\sim$ 2 dex) and the best-fitted
linear regression slope is consistent with 1 within the errors
(0.84$\pm$0.11 for L([OII]) versus L([OIII]) and 1.27$\pm$0.37 for
L([OIII]) versus L([OII]), see Figure~\ref{fig:LOII_LOIII}). The range in
equivalent widths is also similar (1.6 dex for EW([OIII]) and 1.7 dex
for EW([OII])).

If our sample was affected by orientation dependent dust obscuration
(where, as in JB97, substantial numbers of dust clouds lie within the
torus opening angle, and their number increases towards the plane of
the torus) a larger range in L([OIII]) than in L([OII]) would be
observed, due to the obscuration of [OIII] emission at large
inclination angles. Additionally a smaller range in EW([OIII]) than
EW([OII]) would be expected as the result of the orientation dependent
dust reddening of the continuum and the [OIII] emission. If, on the
other hand, only ionization effects were present in our sample, we
would observe a larger range in L([OIII]) than in L([OII]), and a
larger range in EW([OIII)) than EW([OII]), as the [OIII] line is much
more dependent on the ionization parameter $U$ than [OII] (see for
example Simpson 1998 Figure~5). Neither effect is present in our
sample, suggesting that the BQS quasars (at least our radio-quiet
sample) is remarkably free of orientation dependent dust effects or
ionization effects in the narrow-line region (NLR). As both the [OII]
and [OIII] emission are independent of orientation effects, the
[OII]/[OIII] ratio in these optically selected radio-quiet quasars is
not an orientation indicator, contrary to results for radio-loud AGN.

\begin{figure} [t!]
\vspace{9.0truecm}
{\includegraphics{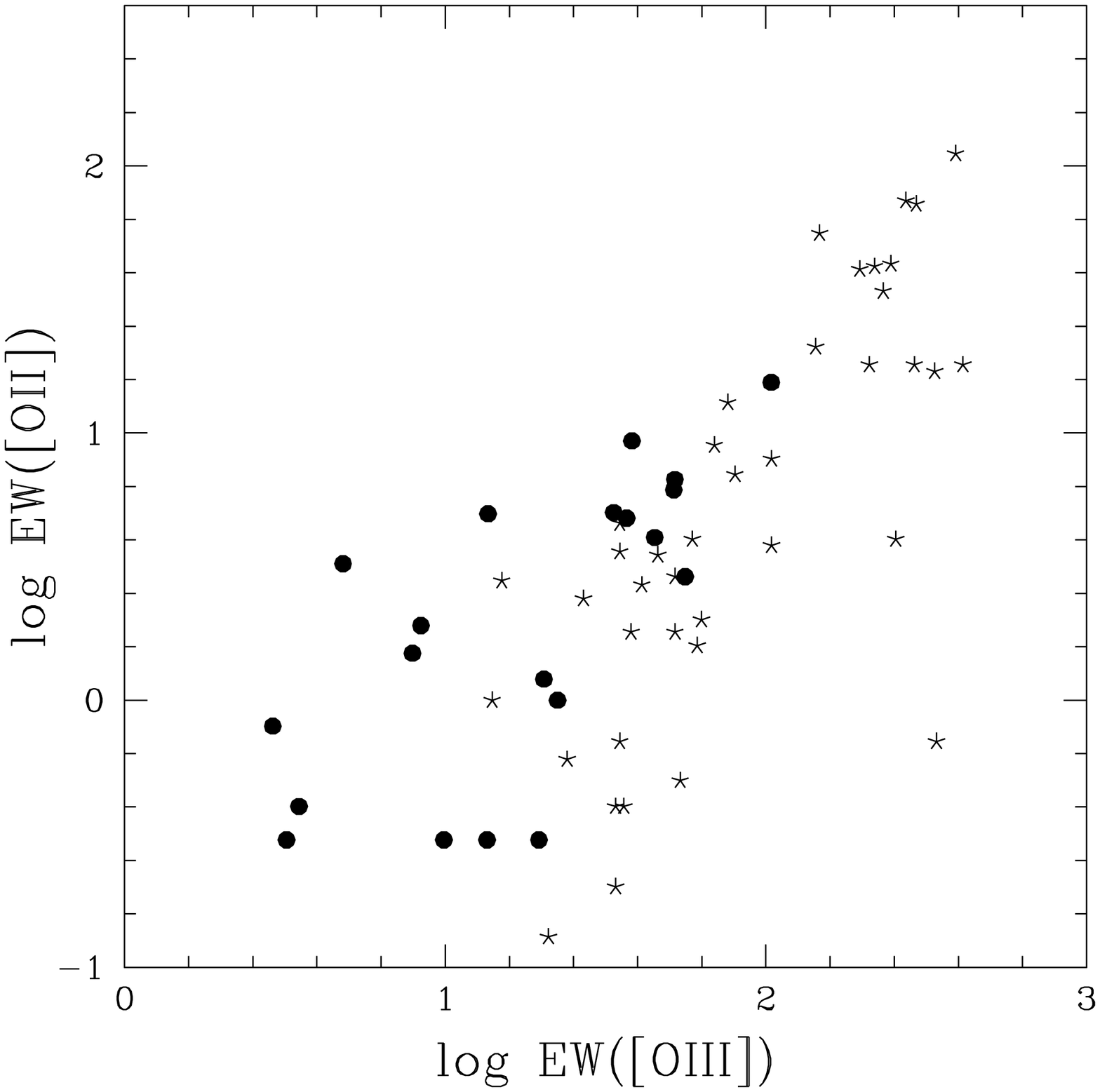}}
{\includegraphics{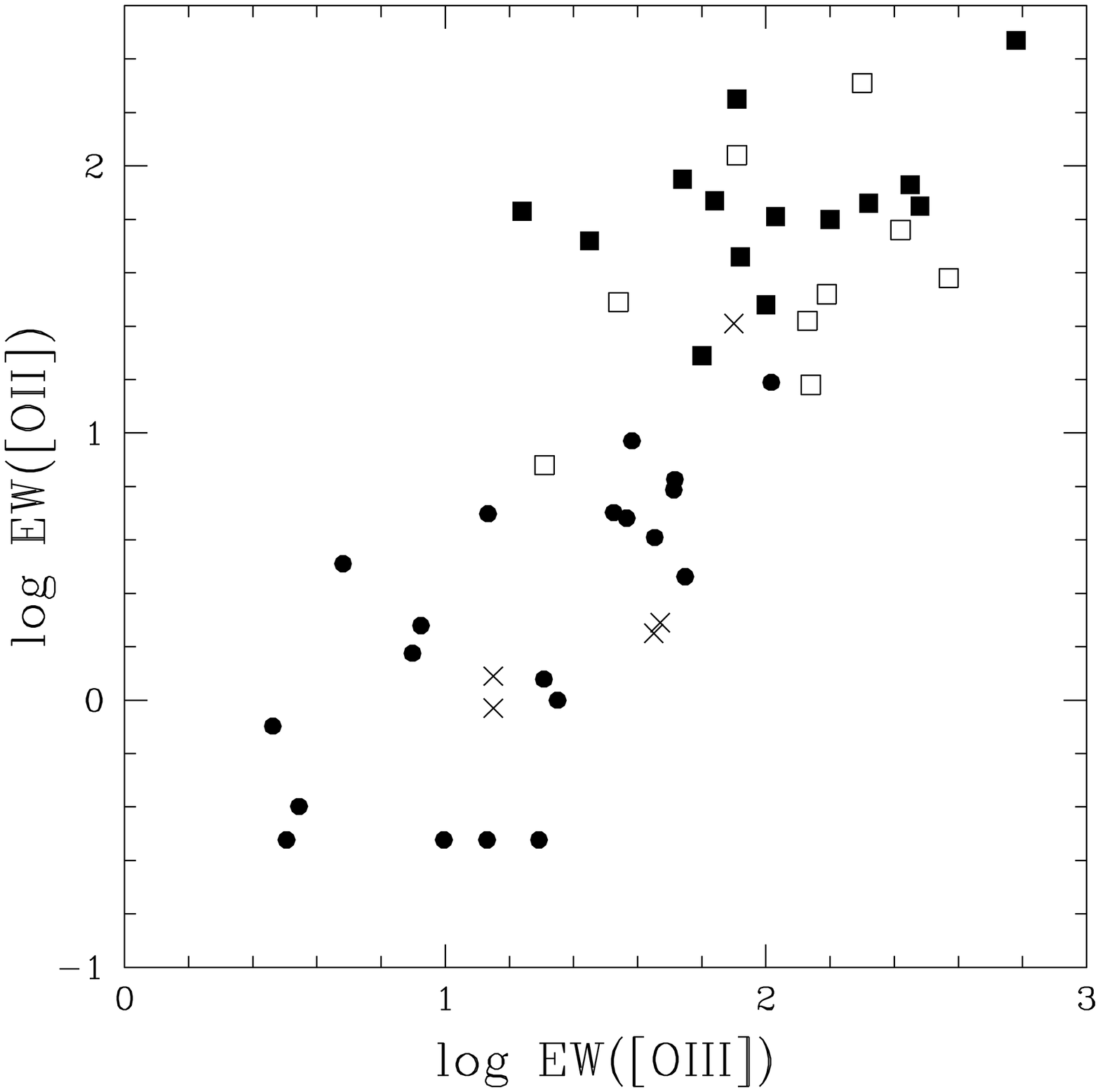}}
\figcaption{a) The comparison of the [OIII] and [OII]
equivalent widths of our sample (filled circles) with Baker (1997;
stars). Our sample extends to smaller EW([OIII]) than JB97 probably
due to our FeII subtraction. The EW([OIII]) can be overestimated by up
to a factor of 10 in strong FeII/weak [OIII] sources (see text) if
FeII is not subtracted. 
b) The comparison of the [OIII] and [OII] equivalent widths of
our sample (filled circles) with Tadhunter et al. (1998) (crosses:
quasars, open squares: broad-line radio galaxies, filled squares:
narrow-line radio galaxies). 
\label{fig:comp_EWOII_OIII}} 
\end{figure}

\begin{figure} [t!]
\vspace{9.0truecm}
{\includegraphics{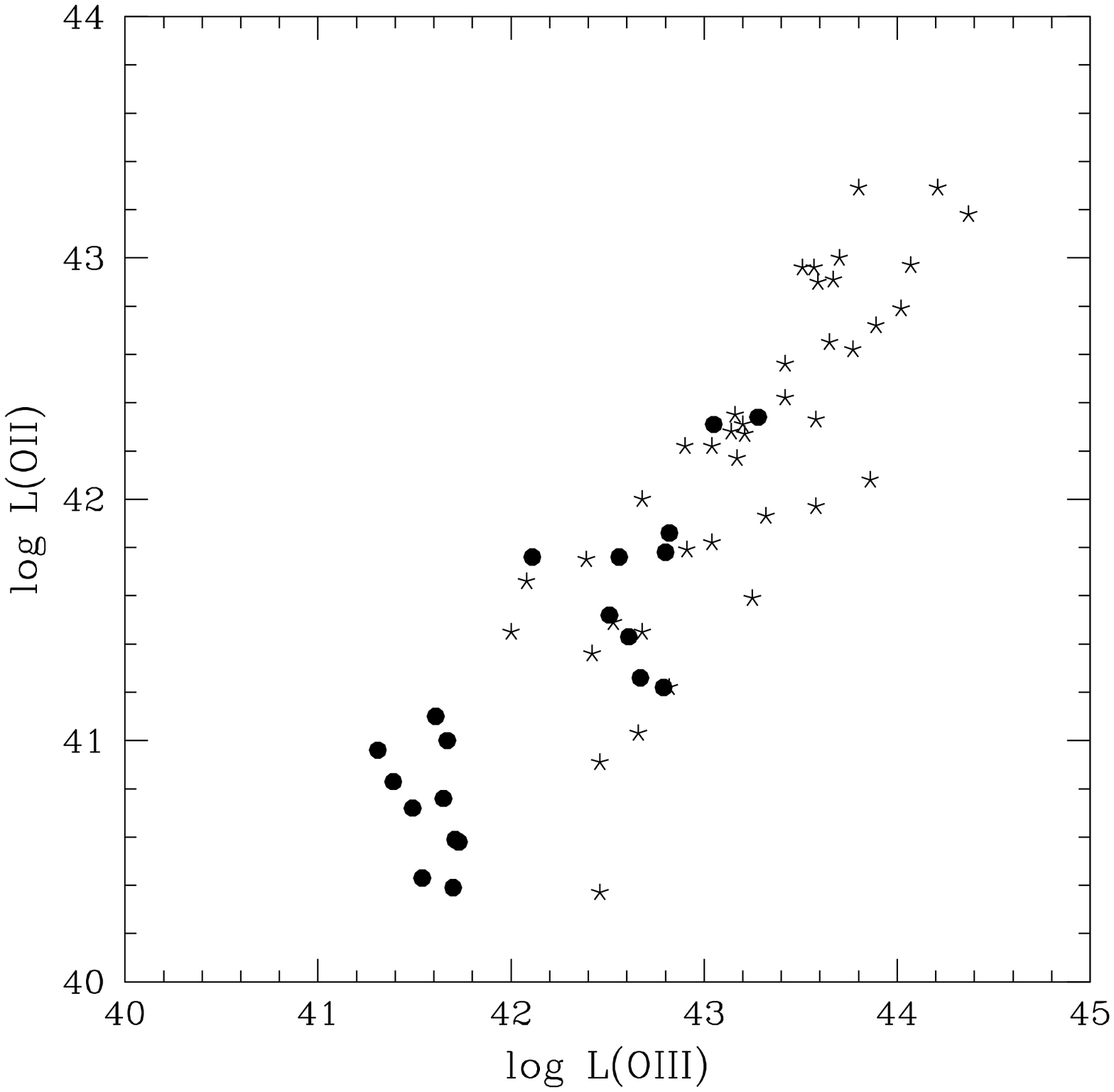}}
{\includegraphics{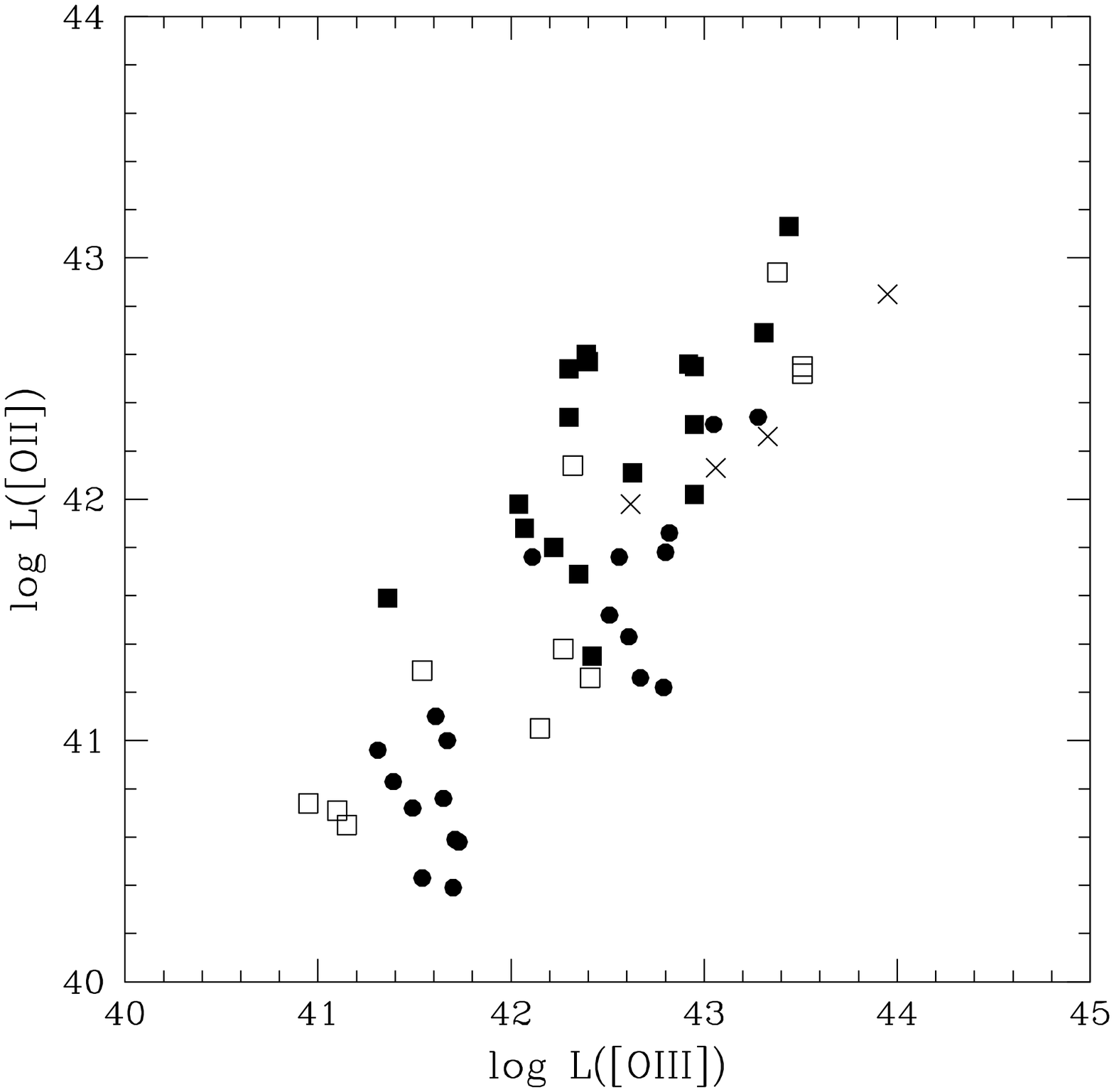}}
\figcaption{a) The comparison of the [OIII] and [OII]
luminosities of our sample (filled circles) with Baker (1997;
stars). b) The comparison of [OIII] and [OII] luminosities of our
sample (filled circles) with Tadhunter et al. (1998; crosses:
quasars, open squares: broad-line radio galaxies, filled squares:
narrow-line radio galaxies). 
\label{fig:comp_LOII_OIII}}
\end{figure}

We compare the equivalent widths and luminosities of [OIII] and [OII]
lines of objects in our optically selected radio-quiet sample with the
JB97 low frequency radio selected quasar sample
(Figure~\ref{fig:comp_EWOII_OIII}a,\ref{fig:comp_LOII_OIII}a) and the
complete sample of southern 2 Jy radio sources presented by Tadhunter
et al. (1998;
Figure~\ref{fig:comp_EWOII_OIII}b,\ref{fig:comp_LOII_OIII}b).  A
number of JB97 quasars and almost all broad- and narrow-line radio
galaxies of Tadhunter et al. are found to occupy a region of higher
EW([OII]) and EW([OIII]) (see Figure~\ref{fig:comp_EWOII_OIII}a,
\ref{fig:comp_EWOII_OIII}b) than our radio-quiet quasars. We found
that these comparison objects cover the whole range of [OIII] and
[OII] luminosities, indicating that the high equivalent widths of the
radio-loud objects are due either to higher [OIII] and [OII]
luminosities or to lower observed continuum. In the latter case the
continuum could be obscured by dust in the radio-selected AGN. This
would confirm previous suggestions (based on the comparison of the BQS
quasars optical slopes [Francis et al. 1991] with the X-ray selected
RIXOS sample of Puchnarewicz et al. 1996 and a heterogeneous sample of
Elvis et al. 1994) that the blue color selection of the BQS QSOs
biases against dust obscured objects while radio-selection is
uneffected.

Figures~\ref{fig:comp_EWOII_OIII}a, \ref{fig:comp_LOII_OIII}a also
show that our sample extends to lower EW([OIII]) and L([OIII]) than
JB97 while having similar cut-off minimum values of EW([OII]) and
L([OII]). The [OII] was measured with respect to the underlying
continuum in both samples and the lowest values are at the detection
limit. It is possible that the [OIII] emission may be overestimated in
the lowest equivalent width/luminosity JB97 objects due to the lack of
FeII subtraction. For the extremely strong FeII objects in our
radio-quiet sample, the equivalent width and luminosity of [OIII] would
be overestimated by a factor of up to 10 if the FeII emission were not
subtracted (e.g. for PG~1402+261 EW([OIII])=36 with FeII included and
EW([OIII])=3 after FeII subtraction).  
Correction for FeII contamination in JB97 sample could potentially
result in an intrinsic range of EW([OIII]) and L([OIII]) larger (by a
factor of 10) than shown in
Figures~\ref{fig:comp_EWOII_OIII}a,\ref{fig:comp_LOII_OIII}a and
comparable to the range of EW([OII]) and L([OII]) respectively. This
would suggest (contrary to the conclusion reached by the author), that
in the JB97 sample there is no dust obscuring the inner region of
[OIII] emission.  Confirmation of this suggestion would require a
re-analysis of the JB97 sample. However to account for the broad line
and continuum reddening observed by JB97, dust between the broad-line
and narrow-line region is still needed.

\subsection{The [OII]/[OIII] ratio as an orientation indicator}

In the previous section we concluded that [OII]/[OIII] ratio is
not an orientation indicator in our radio-quiet BQS sample. In this
section we address the issue of the [OII]/[OIII] ratio as an
orientation measure in radio-loud quasars.  

\begin{figure} [t!]
\vspace{9.0truecm}
{\includegraphics{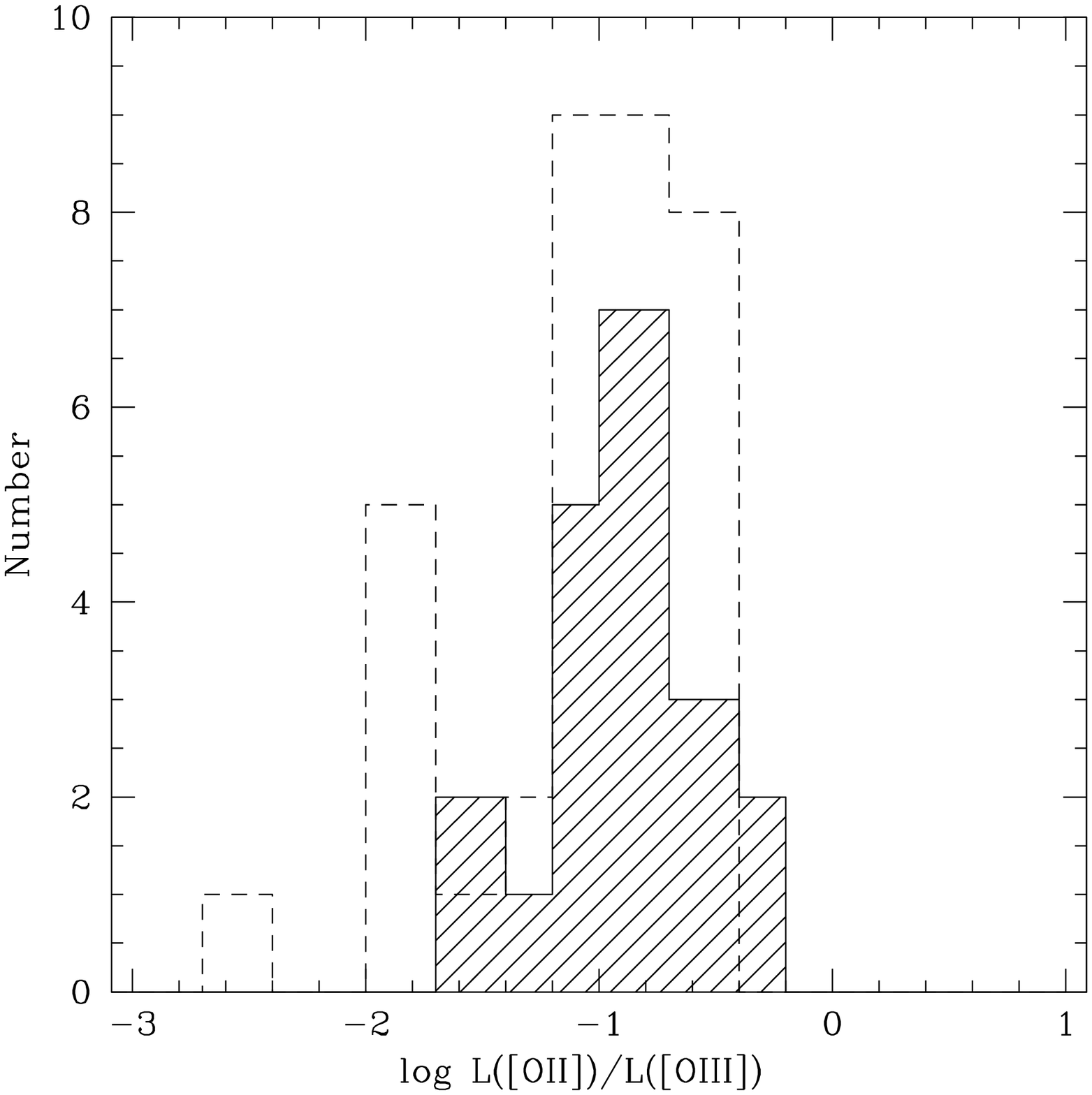}}
{\includegraphics{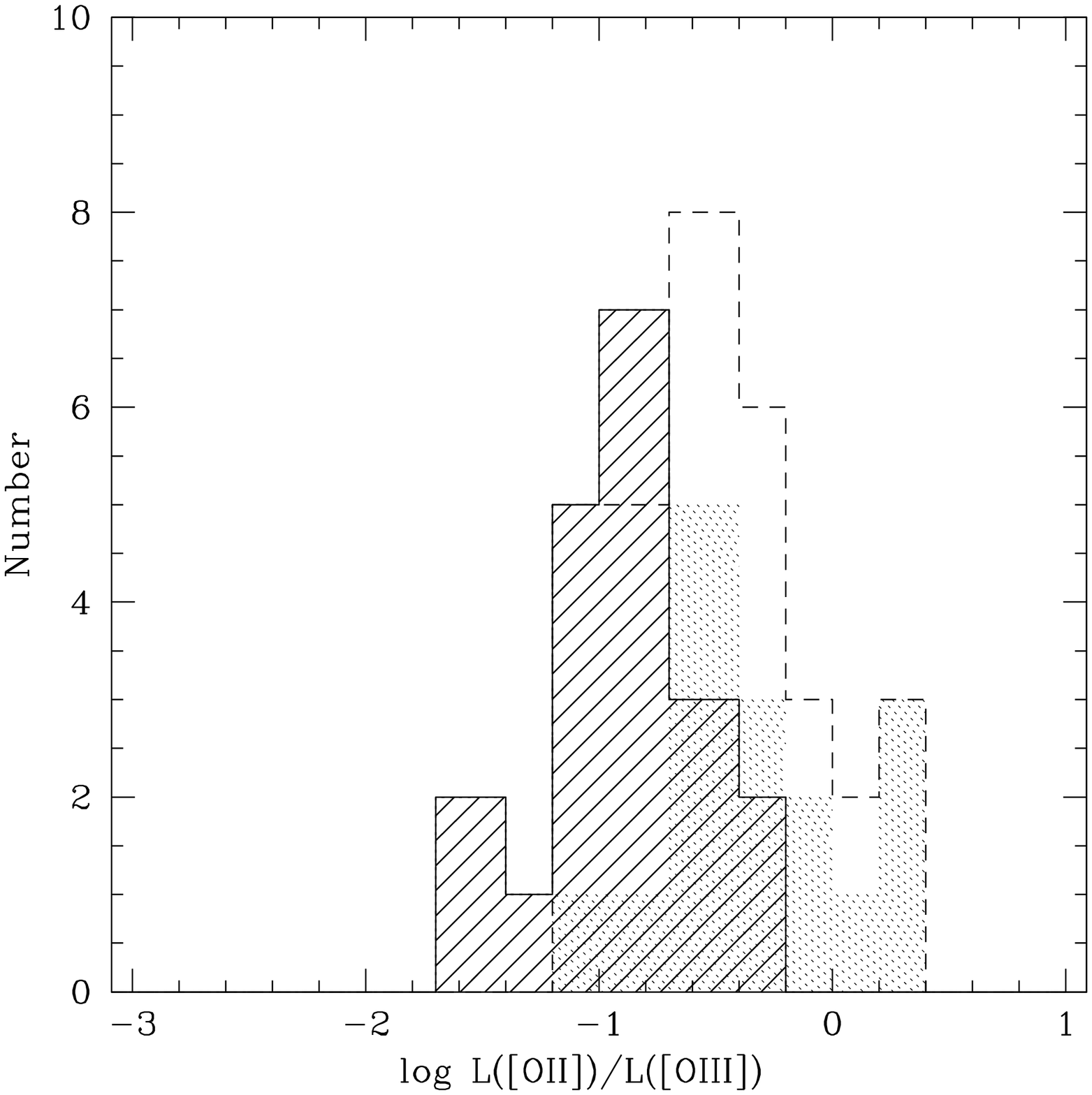}}
\figcaption{a) A histogram of the [OII] to [OIII] luminosity
ratio. Area shaded with solid lines indicates L[OII]/L[OIII] ratios
for our sample, unshaded, dashed contours are radio-loud quasars from
Baker (1997).  b) A histogram of the [OII] to [OIII] luminosity ratio. Area
shaded with solid lines indicates L[OII]/L[OIII] ratios for our
sample, dashed line contours are powerful radio galaxies from Tadhunter et
al. (1998). The area shaded with dotted lines indicates the narrow-line
radio galaxies within the Tadhunter et al. sample. 
\label{fig:OII/OIII}} 
\end{figure}

The comparison of [OII]/[OIII] ratios with JB97
(Figure~\ref{fig:OII/OIII}a) shows a lack of objects in our sample
with the lowest values of [OII]/[OIII] ratio i.e. surprisingly the
most core-dominated objects in JB97. One possibility is an
overestimation of the [OIII] emission in JB97 data resulting from FeII
contamination, as discussed above, which may be the case for four
quasars with the smallest EW([OIII]) (see also Baker et al. 1999 for
spectra).  Core-dominated radio-loud quasars have stronger FeII
emission than lobe-dominated quasars (e.g. Miley \& Miller 1979) so
the FeII contamination would be larger in core-dominated objects
leading to higher apparent [OIII] luminosity (as observed by Jackson
\& Browne 1990) and lower [OII]/[OIII] ratios in JB97. In this case,
the [OII]/[OIII] versus $R$ relation of JB97 could be caused by the
orientation dependence of FeII rather than of [OIII].

Another possible cause of low [OII]/[OIII] ratios in the JB97 sample
(which could be the case for two quasars with extremely large
EW([OIII])) is a higher ionization parameter in radio-loud quasars. 
A comparison of our subset of the BQS sample and the JB97 low
frequency radio selected quasar sample, with the complete sample of 2 Jy
radio sources presented by Tadhunter et al. (1998) shows that both
ours and JB97 samples lack objects with the highest [OII]/[OIII]
ratios ($\log L([OII])/L([OIII]) > 0$ see
Figure~\ref{fig:OII/OIII}b). These high [OII]/[OIII] objects in
Tadhunter et al. (1998) are mostly narrow-line radio galaxies (see
Figure~\ref{fig:OII/OIII}b) believed to be edge-on AGN. These objects
are expected to have a large fraction of the [OIII] nuclear emission
obscured by the dusty torus resulting in a higher [OII]/[OIII] ratio.
However, there are also narrow-line radio galaxies 
(in Figure~\ref{fig:OII/OIII}b) which show values
of L([OII])/L([OIII]) $ < 0$, within the range of ours and the JB97
quasars as well as the broad-line radio galaxies from Tadhunter et
al. This seems to be inconsistent with the orientation dependent
[OIII] scenario in powerful radio-loud galaxies, and suggests that the
[OII]/[OIII] ratio instead depends on the ionization parameter $U$, as 
suggested by Tadhunter et al. (1998). 

Based on our comparisons, we conclude that the [OII]/[OIII] ratio is
not a reliable orientation indicator either in the radio-quiet sample
of the BQS quasars or in the radio-loud quasars.

\section{Conclusions}

Until recently it was generally accepted that eigenvector~1 does not
depend on orientation as it is strongly correlated with [OIII]
emission, originally thought to be an isotropic property in
quasars. As recent studies of radio selected AGN samples have
questioned the isotropy of [OIII] emission, we have investigated the
relation between [OII] emission, which appears to be more isotropic,
and eigenvector~1 and once again addressed the question of orientation
as a driver of eigenvector~1.

We chose radio-quiet quasars from the optically selected Bright
Quasar Survey which showed either high or low [OIII] luminosity,
spanning a wide range of EV1 values in BG92. We subtracted FeII
emission, which contaminates the [OIII] emission, from our spectra
(using the BG92 iron template). We also demonstrated
the significant effect of the presence of the small blue bump (Balmer
continuum and FeII emission) on accurate measurements of the [OII]
emission line, emphasizing the need for spectra covering a wide
($\geq$ 1000\AA) wavelength range in order to determine the underlying
continuum.

We found: 
\begin{enumerate}
\item strong correlations between L([OII]), L([OIII]) and EV1
implying that EV1 does not depend on orientation, confirming 
earlier conclusions of BG92 and Boroson (1992), based on [OIII]
alone. EV1 is likely driven by an intrinsic property
(e.g. accretion rate or black hole spin).  
\item significant EW([OIII])-- EW([OII]) and L([OIII])-- L([OII])
correlations
\item similar ranges in EW([OIII]) and EW([OII]) and in L([OIII]) and
L([OII]) respectively. 
\end{enumerate}
These results lead us to conclude that the optically selected BQS
sample (at least our radio-quiet sample) is free from orientation
dependent dust effects and ionization dependent effects in the
narrow-line region. Assuming our sample is representative of bright,
optically selected radio-quiet quasars, this implies that their [OIII]
emission is isotropic and the [OII]/[OIII] ratio is not an orientation
indicator. This is in contrast with earlier results for the radio
selected AGN (Baker 1997; Jackson \& Browne 1990). We suggest that
this discrepancy may be due to, contamination of the [OIII] emission
by orientation dependent FeII emission in the latter samples.

Acknowledgements - We are grateful to Perry Berlind for observing the
spectra of our sample quasars, Martin Elvis and Joanne Baker for
helpful discussions, and Todd Boroson for providing the FeII optical
template and the eigenvector~1 values. We gratefully acknowledges the
support: of the Smithsonian pre-doctoral fellowship at the
Harvard-Smithsonian Center for Astrophysics and grant no. 2P03D00410
of the Polish State Committee for Scientific Research (JK), NASA
contract NAS8-39073(CXC) (BJW), NASA LTSA grant NAG5-8107 and the
Alfred P. Sloan Foundation (WNB), and a Research Assistantship at SAO
made possible through NASA grants: NAGW-4266, NAGW-3134, NAG5-4089 to
BJW and the Columbus Fellowship at The Ohio State University (MV).

%
%
%

\begin{figure} [t!]
\vspace{9.0truecm}
{\includegraphics{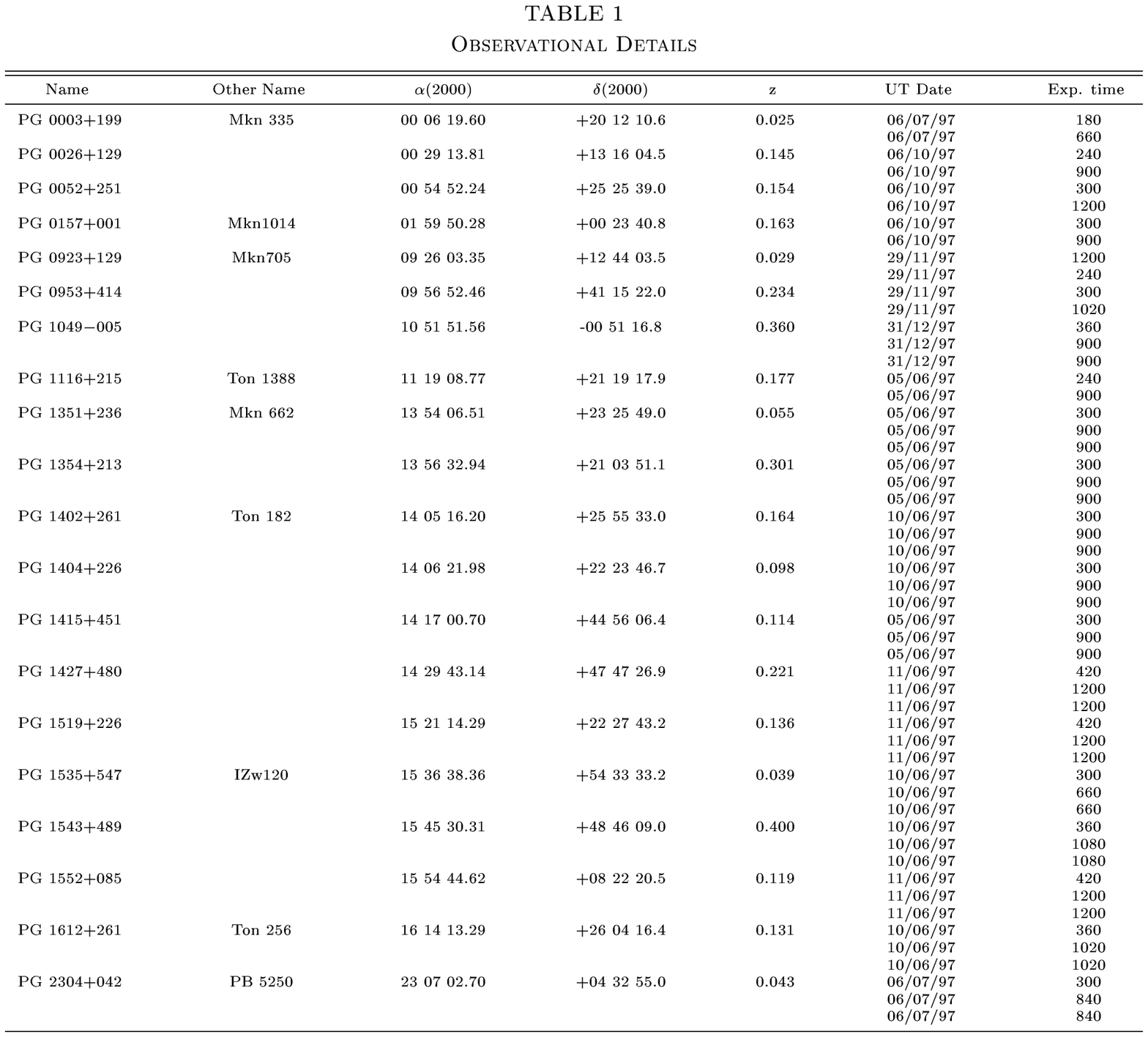}}
\end{figure}

\clearpage

\begin{figure} [t!]
\vspace{9.0truecm}
{\includegraphics{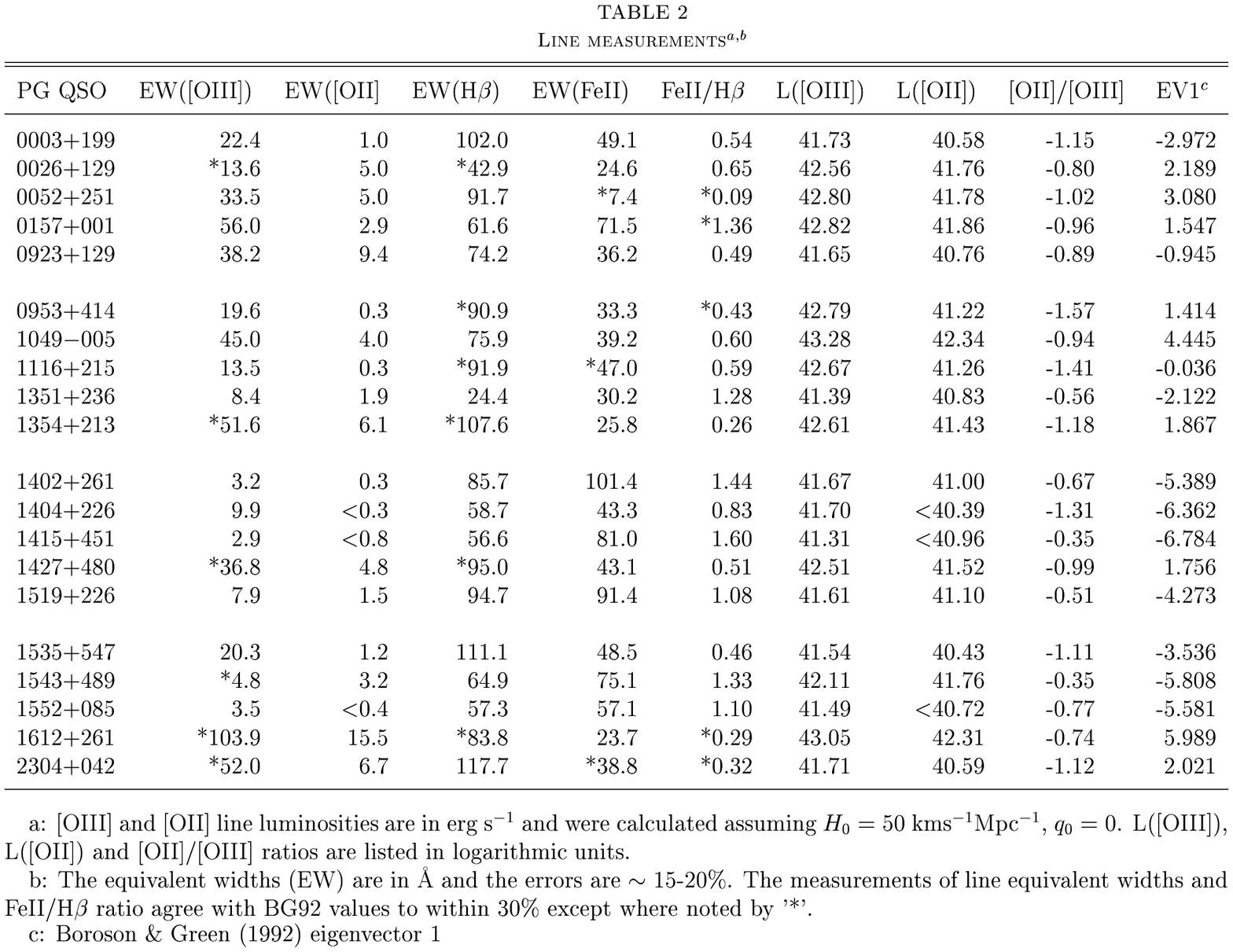}}
\end{figure}

\end{document}